\documentclass{article}

\usepackage[a4paper, total={6in, 8in}]{geometry}

\usepackage{graphicx} 
\usepackage{xcolor} 
\usepackage{subcaption} 
\usepackage{amsmath,bbm} 
\usepackage{amssymb} 
\usepackage{hyphenat} 
\usepackage{authblk} 
\usepackage{hyperref} 
\usepackage{comment} 
\usepackage[most]{tcolorbox}
\usepackage{xurl}
\usepackage{float}
\usepackage[noabbrev,capitalise]{cleveref}
\usepackage{algorithm}
\usepackage{algpseudocode}

\title{Automated Repair of AI Code with Large Language Models and Formal Verification}
\author[1]{Yiannis Charalambous}
\author[1]{Edoardo Manino}
\author[1,2]{\authorcr Lucas C. Cordeiro}
\affil[1]{The University of Manchester, Department of Computer Science, Manchester, United Kingdom}
\affil[2]{Federal University of Amazonas, Manaus, Brazil}
\date{}

\begin{document}

\maketitle

\begin{abstract}
    The next generation of AI systems requires strong safety guarantees. This report looks at the software implementation of neural networks and related memory safety properties, including NULL pointer deference, out-of-bound access, double-free, and memory leaks. Our goal is to detect these vulnerabilities, and automatically repair them with the help of large language models. To this end, we first expand the size of \textit{NeuroCodeBench}, an existing dataset of neural network code, to about $81$k programs via an automated process of program mutation. Then, we verify the memory safety of the mutated neural network implementations with ESBMC, a state-of-the-art software verifier. Whenever ESBMC spots a vulnerability, we invoke a large language model to repair the source code. For the latest task, we compare the performance of various state-of-the-art prompt engineering techniques, and an iterative approach that repeatedly calls the large language model.
\end{abstract}

\section{Introduction}
\label{sec:intro}

In contrast to classic software development, neural networks are crafted via a long process of trial and error that terminates when their predictive performance reaches a satisfactory level~\cite{Biderman2020,Suresh2021}. The iterative and performance-driven nature of this process leaves neural networks vulnerable on many fronts~\cite{Huang2020}: poor performance on out-of-distribution~\cite{Fort2021} and adversarial inputs~\cite{Narodytska2017}, misspecification of the neural architecture and training process~\cite{Humbatova2020}, invocation of broken and deprecated libraries~\cite{Morovati2023}, outright software bugs~\cite{Guo2021}. Unfortunately, many of these vulnerabilities are not easy to catch early in the development process and may remain hidden until after deployment.

Although efforts to debug the actual implementation of neural networks exist, they are based on automatic testing and thus cannot prove correctness for all inputs~\cite{Odena2019,Guo2021,Deng2023}. This lack of guarantees is especially concerning for safety-critical systems since common software vulnerabilities~\cite{cwe} (e.g., arithmetic overflows, invalid memory accesses) can make the networks produce wrong results, expose sensitive data or corrupt the system they are executed on.

In this report, we tackle the challenge of producing bug-free implementations of neural networks in the following way. First, we employ software verifiers to ensure full coverage of the state space. In the past, it has been claimed that software verifiers struggle to cope with neural network code due to its size, complexity and the presence of numerous calls to the standard mathematical library~\cite{manino2023neurocodebenchpub}. To verify this claim, we generate a large dataset of neural network code, by injecting memory vulnerabilities via program mutations. Our results show that software verifiers can find memory safety vulnerabilities in neural network code relatively easily, thus allowing us to use them as oracles for the correctness of neural network implementations.

Second, we adopt large language model as a powerful tool for program repair. In the past years, large language models have shown great promise in a wide variety of code processing tasks including program translation \cite{white2023prompt}, code completion \cite{liu2016neural} and automated repair \cite{charalambous2023new}. At the same time, the related literature always show a drop in performance on out-of-distribution samples, which are different from those seen during training. Here, we posit that neural network implementations, or AI code in short, fits in the latter category and constitutes an excellent benchmark for out-of-distribution performance. Indeed, our results show that off-the-shelf large language models can repair AI code, but require very specific prompting techniques to do so at an acceptable level.

More specifically, this report covers the following topics:
\begin{itemize}
    \item Our methodology to generate a large dataset of AI code examples. This is done by automatically increasing the size of the \textit{NeuroCodeBench} dataset~\cite{manino2023neurocodebenchpub} via code-specific data augmentation techniques.
    \item Experimental results on running the ESBMC software verifiers on the augmented dataset to classify benchmarks that contain memory safety vulnerabilities.
    \item A comparison of different prompt engineering techniques to optimise the repair performance of large language models. In this respect, we propose a few solutions to the issues of limited context windows, code formatting and including feedback from compilers and software verifiers.
    \item Preliminary results on the repair performance of large language model. Specifically, we analyse their output along four axes: correct syntax, relevance to task, compilability and successful repairs.
    \item Comparing and finding the best format of the history in iterative code repair affects iterative code repair performance.
    \item Impact of temperature on iterative code repair.
\end{itemize}

The structure of this report is the following. In Section~\ref{sec:background}, we give some background on \textit{NeuroCodeBench}, common safety vulnerabilities found in neural networks and the program mutation techniques we employ. In Section \ref{sec:expanded_dataset}, we detail our methodology towards creating an AI code dataset, we present our empirical results in using ESBMC to label the dataset. In Section \ref{sec:llm_prompting} we introduce a wide variety of state-of-the-art prompting techniques, we show how to employ them for code repair, and compare their code repair performance with ChatGPT.

This report is to be considered as the official documentation of the public software repository at \url{https://github.com/emanino/plain_c_nn_benchmark}, the staging repository can be found at \url{https://github.com/Yiannis128/plain_c_nn_benchmark}.

\section{Background}
\label{sec:background}

\subsection{NeuroCodeBench}
\label{sec:background_neurocodebench}

\begin{table}[t]
\centering
\resizebox{\textwidth}{!}{%
\begin{tabular}{ |c|c|c|c|c|c|c|c| } 
    \hline
    Neural Network Category & Inputs & Outputs & Layers & Neurons & Activations & Architecture & Conversion \\
    \hline
    \texttt{hopfield\_nets} & 4--64 & 4--64 & 1 & 4--64 & Softsign/TanH & Recurrent & \texttt{keras2c} \\
    \texttt{poly\_approx} & 1 & 1 & 1--4 & 16--1024 & ReLU & Feedforward & \texttt{keras2c} \\
    \texttt{reach\_prob\_density} & 3--14 & 3--14 & 2--3 & 64--192 & ReLU & Feedforward & \texttt{onnx2c} \\
    \texttt{reinforcement\_learning} & 4--8 & 2--8 & 2 & 128--512 & ReLU & Feedforward & \texttt{onnx2c} \\
    \hline
\end{tabular}}
\caption{Neural networks in \textit{NeuroCodeBench}. The ``Layers'' and ``Neurons'' columns refer to the hidden layers only. The networks in \texttt{hopfield\_nets} have a number of iterations between $1$ and $4$.}
\label{tab:network_overview}
\end{table}

\textit{NeuroCodeBench} is a plain C benchmark of neural network implementations designed for formal software verification. In general, mainstream machine learning libraries (e.g., PyTorch~\cite{PyTorch} and Tensorflow~\cite{TensorFlow}) have an opaque multi-language interpreted structure that can be easily tested~\cite{Guo2021,Deng2023}, but does not lend itself to automated software verification. For this reason, \textit{NeuroCodeBench} opts for micro-controller frameworks, where the network's source code is fully available. More specifically, it uses two existing tools for converting high-level neural network specifications to standalone C code: \texttt{onnx2c}\cite{onnx2c} and \texttt{keras2c}\cite{keras2c,Conlin2021}. Since November 2023, \textit{NeuroCodeBench} is part of the official benchmark for the international software verification competition (SV-COMP).\footnote{\url{https://gitlab.com/sosy-lab/benchmarking/sv-benchmarks/-/merge_requests/1456}}

In Table \ref{tab:network_overview}, we give an overview of the neural architectures of \textit{NeuroCodeBench}. These have been designed to cover several use cases in machine learning and engineering, as detailed below:

\paragraph{Hopfield Networks.}

For a long time, it has been known that certain types of recurrent neural networks can act as error-correcting decoders~\cite{AbuMostafa1985,Chaudhuri2019}. The main idea is encoding a sequence of $d$ bits into a vector $x\in\{\pm1\}^d$, and letting the neural network flip the sign of the incorrect entries. Here, we choose Hopfield networks with Hebbian weights since their properties are well understood~\cite{AbuMostafa1985}. Specifically, we build networks reconstructing a single pattern $x=(1,\dots,1)$. We vary the pattern length in $d\in\{4,8,16,32,64\}$ and the number of recursions in $t\in[1,4]$. For compatibility with \texttt{keras2c}~\cite{keras2c,Conlin2021}, we use Softsign and TanH activations (see Table~\ref{tab:network_overview}) rather than traditional sign activations~\cite{AbuMostafa1985}.

\paragraph{Polynomial Approximation Networks.}

In several engineering areas, neural networks are used to approximate the transfer function of electrical components~\cite{Xu2002,Massi2023}. We emulate this process by defining a hypothetical polynomial component $f(x) = 0.125 x^4 - 0.25 x^3 - 0.75 x^2 + x + 0.5$ with an oscillating transfer function. Then, we create a training set by uniformly sampling $f(x)$ in $x\in[-2,3]$ and train $16$ different feedforward ReLU networks $\hat{f}(x)$. The smallest has four layers with four neurons each, and the largest has a single hidden layer with $1024$ neurons (see \texttt{poly\_approx} category in Table~\ref{tab:network_overview}).

\paragraph{VNN-COMP Networks.}

Since its first edition in $2020$, the International Verification of Neural Networks Competition (VNN-COMP) has published all its benchmarks~\cite{Mueller2023}. These benchmarks do not contain implementation details since they target a higher level of abstraction (real number arithmetic, no memory model). To provide a reference implementation, we translate the networks from ONNX format~\cite{ONNX} to C with \texttt{onnx2c}~\cite{onnx2c}. From the 2022 edition~\cite{Mueller2023}, we select two categories with small networks (see Table \ref{tab:network_overview}): \texttt{reach\_prob\_density}~\cite{Meng2022} and \texttt{reinforcement\_learning}~\cite{Ravaioli2022}.

\subsection{Safety Properties}
\label{sec:background_safety_prop}

Originally, \textit{NeuroCodeBench}~\cite{manino2023neurocodebenchpub} was designed with the purpose of testing software verifiers. As such, it features very challenging verification problems centered around reachability properties~\cite{daws1998reachability}. In contrast, our AI code dataset is meant to test the ability of large language models to repair code. As the majority of real-world bugs~\cite{Humbatova2020} are the result of programming mistakes, in this report, we focus on memory safety properties instead. Indeed, memory safety is the leading cause of software vulnerabilities~\cite{miller2019trends}. Here, we include a brief description of the different safety properties.

\paragraph{Reachability.} For neural networks, we define reachability properties as necessary conditions over a set of inputs $x\in\mathcal{X}$ and outputs $y\in\mathcal{Y}$, where $y=f(x)$ and $f$ is the neural network \cite{brix2023fourth}. In general, a reachability property would take the following form:
\begin{equation}
\label{eq:nn_reach}
    \forall x\in \mathcal{X}, y=f(x)\implies y\in \mathcal{Y}
\end{equation}
As a concrete example, consider the following. Take an error-correcting Hopfield network, which is trained to reconstruct the code $y=(1,1,\dots,1)$ in the presence of input noise. Assume that all input vectors $x$ of dimension $d$ have at most three flipped inputs, i.e., $\mathcal{X}=\{x|x_i\in\{\pm1\}\land\sum_i^dx_i\geq d-6\}$. The neural network is always able to reconstruct the correct code if Equation~\ref{eq:nn_reach} holds for $\mathcal{Y}=\{y=(1,1\dots1)\}$.

\paragraph{Memory safety.} Properties encompass checking vulnerabilities regarding NULL pointer dereferences, out-of-bounds accesses, double-free, and memory leaks. In more broad terms, we check for situations where an illegal memory read or write can occur. ESBMC can check these properties by default by enabling the correct flags. For running the experiments, the following flags were enabled: 
\texttt{--interval-analysis --goto-unwind --unlimited-goto-unwind --incremental-bmc \\ 
--state-hashing --add-symex-value-sets --k-step 2 --floatbv --unlimited-k-steps \\ 
--memory-leak-check --context-bound 2 --timeout 300 -Iincludes -Inetworks}. \\
These properties instruct ESBMC to encode memory safety criteria within the symbolic execution engine, ensuring that (1) all memory deallocations are valid, (2) all pointer dereferences are valid, and (3) all allocated memory is accurately tracked, encompassing pointed to or deallocated.

\subsection{Generative Language Models}

\begin{figure}[h]
    \centering
    \includegraphics[width=0.5\textwidth]{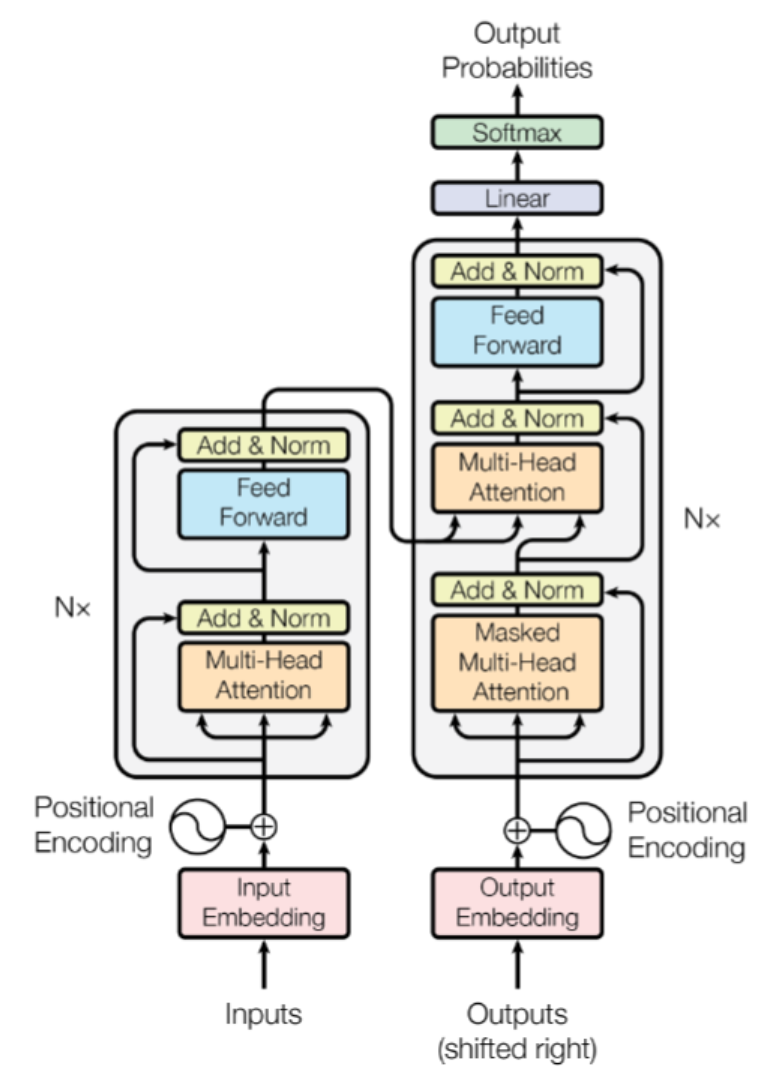}
    \caption{Diagrammatic layout of a transformer model architecture \cite{vaswani2017attention}.}
    \label{fig:attention}
\end{figure}

AI has been used as mutation generators for automatically repairing code. In recent years, Large Language Models (LLMs) have taken precedence \cite{fan2023automated}, however, the code that they generate is not secure and is filled with vulnerabilities \cite{charalambous2023new}. LLMs are a type of Recurrent Neural Network (RNN), an AI architecture that uses an encoder/decoder architecture, typically along with an attention mechanism \cite{vaswani2017attention}. Opposite to conventional Recurrent Neural Networks (RNNs) that compute their values over a sequence of hidden states, LLMs with the encoder/decoder architecture can be heavily parallelized, allowing for more training in less time \cite{vaswani2017attention}. \cref{fig:attention} shows a diagram of the layout of LLMs.

Furthermore, these developments have paved the way for OpenAI's \textit{GPT-3.5-Turbo}, a closed source LLM. GPT-3.5-Turbo is a 175 billion parameter autoregressive language model \cite{brown2020language} that will be used in the experiments in this report.

\subsection{Software Verification}

\begin{figure}
    \centering
    \includegraphics[width=0.6\textwidth]{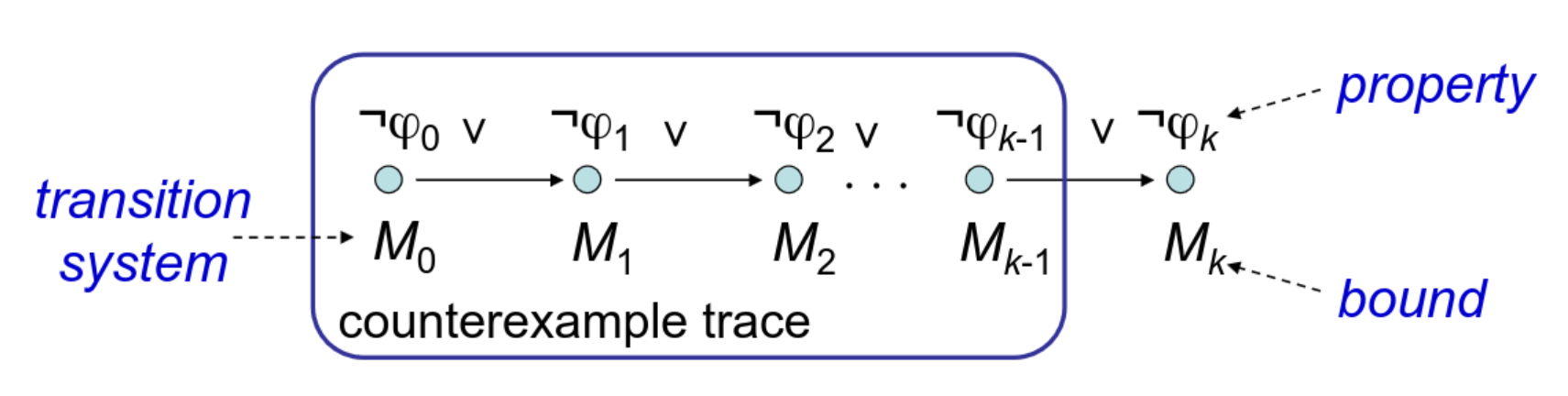}
    \caption{A diagram visualization of how bounded model checking works. The visualization represents a state transition system $M$, a given property $\phi$, and a verification condition $\psi$ \cite{lucassoftwaresecurity}.}
    \label{fig:bmc}
\end{figure}

There are many methods of verifying software, in this report we use Bounded Model Checking (BMC) to detect if the code in the experiments contains any memory vulnerabilities \cite{cordeiro_smt-based_2009}. BMC is a technique used where a given property is checked at a specified depth in a system \cite{cordeiro_smt-based_2009}. To elaborate, given a state transition system, and a property, BMC unrolls the system and translates it to a verification condition, if the verification condition is true, then the system is satisfiable up to that specified bound \cite{cordeiro_smt-based_2009}. \cref{fig:bmc} visualizes how BMC works. The Efficient SMT-based Bounded Model Checker (ESBMC) is such a software verifier that uses bounded model checking to prove mathematically up to a certain depth the presence of software vulnerabilities \cite{cordeiro_smt-based_2009}. This report will use ESBMC to test samples of AI C code for the presence of vulnerabilities.

\section{Creating An AI Code Dataset}
\label{sec:expanded_dataset}

In order to fine-tune and evaluate the performance of large language models in repairing AI code, we need a dataset of neural network implementations. Since existing ones are fairly small, amounting to only a few hundred samples~\cite{manino2023neurocodebenchpub}, we turn to data augmentation techniques to greatly expand their size. This is needed to obtain statistically significant evaluation metrics.

\subsection{Program Mutation}
\label{sec:background_mutation}

The test suite chosen for this task is the ``practical mutation testing tool for C and C++'' Mull~\cite{8411727}. Mull comprises two key components, the compiler plugin and Mull Runner, an entirely separate program~\cite{8411727}. The compiler plugin generates program mutations at the compilation stage and injects them into the LLVM bit code~\cite{8411727}. The mutations enabled in Mull's configuration files are injected into the IR of the program but are hidden behind conditional flags that allow them to be individually enabled during runtime~\cite{8411727}.

Mull Runner can run the program repeatedly with each mutation condition enabled~\cite{8411727}. Mull runner will then check how the injection of the mutation affects the program's execution~\cite{8411727}. Each mutation can be saved into a patch file and stored on disk~\cite{8411727}.

\subsection{Data Augmentation}
\label{sec:data_augmentation}

\begin{figure}[h]
\centering
    \includegraphics[width=1.0\textwidth]{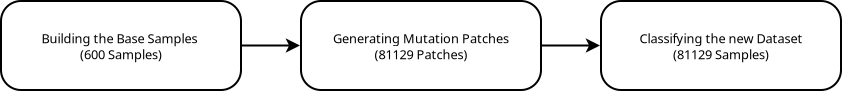}
\caption{Overview of the \textit{NeuroCodeBench} expansion pipeline. It consists of three key stages.}
\label{fig:expansion_methodology}
\end{figure}

The \textit{NeuroCodeBench} dataset~\cite{manino2023neurocodebenchpub} was expanded by a pipeline that aimed to automate the dataset generation as much as possible. The sheer size of the dataset is the reason for this, as it is not realistic to create a dataset with $81$k samples in such a short time span using manual labor methods. The pipeline can be divided into three distinct sections that process and transform the base dataset from one form to the next. The process begins by building the base sample source code, followed by generating mutation patches using a mutation test suite; finally, the last step consists of using ESBMC to classify the samples. Figure~\ref{fig:expansion_methodology} illustrates an overview of this pipeline process. 

\paragraph{Building the Base Samples.} We follow the same procedure outlined by the \textit{NeuroCodeBench} instructions for building the base dataset. The base dataset contains $505$ samples. There are $79$ Hopfield networks, $97$ polynomial approximation networks, $34$ probability density networks, and $295$ reinforcement learning networks.

\paragraph{Generating Mutation Patches.} Having obtained the base dataset, we use a mutation testing suite to generate patches for each sample that modify the code in a small way (see Section~\ref{sec:background_mutation}). 

While Mull is designed to evaluate the quality of test suites for C programs~\cite{8411727}, we have managed to use it to multiply the size of the base dataset by around $80$x, reaching $81129$ total samples. This is because we use the diff tool $d$ to apply each patch generated $y_x$ to its original file $f$, yielding a new transformed file $f'$ in the following process $f'=d(f, y_x)$. Patches are produced through the Mull runner program \$m\$ via the following procedure: \$y=m(f, c)\$, where \$c\$ denotes Mull's configuration containing the specified mutations allowed for patch creation, and \$y\$ represents the collection of patches generated by the Mull compiler plugin. Creating a new sample for every patch expands the size of the samples from $505$ to $81129$ samples; this increase depends on the amount of mutations that are enabled in the Mull configuration.

\textit{NeuroCodeBench} samples also include calls \texttt{\_\_VERIFIER\_assume(expression)} and \texttt{\_\_VERIFIER\_assert(expression)} which instruct the verifiers to verify certain safety properties, as discussed in Section~\ref{sec:background_safety_prop}. As discussed previously, removing those statements is necessary as this dataset focuses on memory safety only\footnote{Note that the expanded sample dataset may contain some duplicates after the assert declarations are removed.}.

\paragraph{Classifying the new Dataset.} With the expanded dataset generated, ESBMC can then be used to verify the safety of each sample. As we are verifying only memory safety properties, the number of samples ESBMC can classify should be higher than when verifying the additional safety properties of the base dataset. 

\subsection{Empirical Results}
\label{sec:results}

This section explores the experimental setup and results of processing this extended \textit{NeuroCodeBench} dataset using the ESBMC verifier. The experiments were conducted on 6 machines with Intel(R) Xeon(R) CPU E5-2620 v4 @ 2.10GHz, 198 GB RAM.

\begin{figure}[t]
\centering
    \begin{subfigure}[t]{0.59\textwidth}
    \centering
        \includegraphics[height=3.2cm]{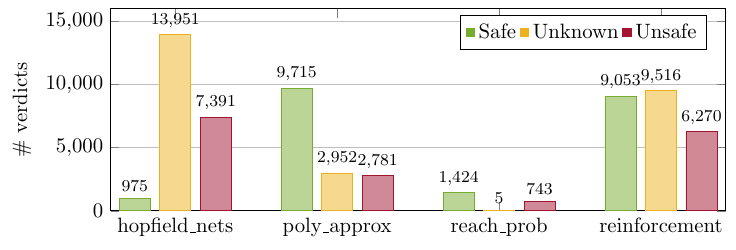}
        \caption{Number of safe and unsafe mutated programs. If ESBMC times out after 300s, the program is labeled unknown.}
        \label{fig:esbmc_verdict}
    \end{subfigure}
    \hfill
    \begin{subfigure}[t]{0.39\textwidth}
    \centering
        \includegraphics[height=3.2cm]{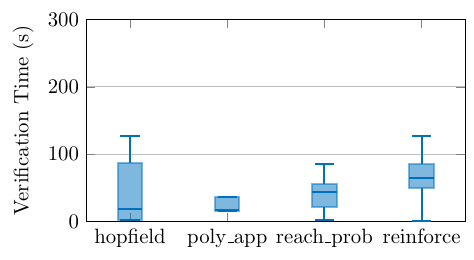}
        \caption{Verification time for a small subset of safe and unsafe programs.}
        \label{fig:esbmc_time}
    \end{subfigure}
\caption{Performance of ESBMC on $25952$/$81129$ mutated programs, by neural network category.}
\label{fig:results}
\end{figure}

\paragraph{Classification Results} Figure~\ref{fig:esbmc_verdict} shows classifications assigned by ESBMC to each sample in each category. Note that each category exhibits different ratios of safe, unsafe, and unknown verdicts. Hopfield networks have the smallest amount of safe cases, compared to the amount of unsafe and unknown cases that they have, this is possibly due to the complexity of Hopfield networks, as they are recurrent neural networks. Probability density and reinforcement learning networks have a much higher portion of safe and unsafe cases compared to unknowns. These observations suggest that ESBMC is more capable at processing the latter two network types, possibly due to their simpler structure, as the symbolic execution would be less demanding due to possibly having less loop unrolling.

\paragraph{Verification Time} Figure \ref{fig:esbmc_time} shows a box-plot of each category, illustrating various information about the verification time for safe or unsafe cases. As can be observed, Hopfield and polynomial approximation neural networks are much faster to complete on average than probability density and reinforcement learning graphs. One possible explanation is that these two categories of benchmarks use a different neural network library (keras2c) than the others (onnx2c). However, more data is needed to get a conlusive answer. Finally, note that while ESBMC could not successfully classify various samples, a sizable portion was processed successfully within a range of a few minutes.

\subsection{Lessons Learned}
\label{sec:future}

Overall, ESBMC is relatively efficient in verifying neural network code containing memory safety vulnerabilities. This contrasts with previous work on \textit{NeuroCodeBench}, which showed that software verifiers such as ESBMC struggle to reason over reachability properties of neural networks. Still, the moderately large ratio of unknown verdicts, together with the long timeout of $300$ seconds (or $5$ minutes), leave room for improvements in terms of verification time. In the future, we might consider a portfolio approach combining falsification tools (e.g., fuzzers) with verification tools such as ESBMC.

For the time being, the sizeable number of conclusive verdicts (safe or unsafe) still allows us to build a large AI code dataset. With this dataset, we can explore the effectiveness of LLMs in recognizing potential vulnerabilities and repairing unsafe AI programs.

\section{Repairing AI Code with Large Language Models}
\label{sec:llm_prompting}

The AI code dataset we generate in Section~\ref{sec:expanded_dataset} is a representative instance of \textit{out-of-distribution} data. Indeed, while the original \textit{NeuroCodeBench} is in the public domain, it was released in late 2023~\cite{manino2023neurocodebenchpub}, and it is thus unlikely to have been included in the training set of most state-of-the-art large language models~\cite{radford2019language}. In addition, our automated mutation technique further ensures that our AI code dataset looks very different from any piece of software the current large language models have been trained on.

\begin{figure}
    \centering
    \includegraphics[width=0.8\textwidth]{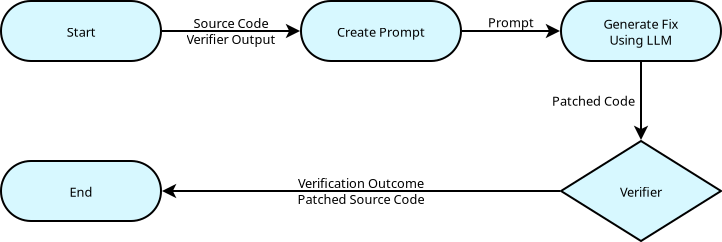}
    \caption{Flowchart of a single attempt at repairing AI C Code with the LLM.}
    \label{fig:singleattemptrepairdiagram}
\end{figure}

Against this background, we pose the following research question: \textit{can off-the-shelf large language models spot the memory vulnerabilities we introduced in Section~\ref{sec:expanded_dataset} and eliminate them by repairing the code?} \cref{fig:singleattemptrepairdiagram} displays a visualization of the question.

\subsection{Prompt Templates}
\label{sec:prompt_engineering}

The consensus in the natural language processing community is that the performance of large language models depends on our ability to prompt them effectively~\cite{white2023prompt}. This has sparked much interest in \textit{prompt engineering} techniques, which have sometimes entirely replaced the need to finetune a large language model~\cite{white2023prompt}. Here, we list a few state-of-the-art prompt engineering strategies relevant to the code generation task. Our overarching goal is to compare their empirical performance (See Section~\ref{sec:prompt_comparison}).

\subsubsection{Previous Experience with ESBMC-AI}
\label{sec:prompt_old}

An earlier technical report~\cite{charalambous2023new} showed promising results by prompting large language models to repair vulnerable code. We included their prompt template in our experimental evaluation to establish a baseline. We list it here, with appropriate placeholders for the code to be repaired (\texttt{\{source\}}) and the verifier feedback (\texttt{\{esbmc\}}). We discuss dealing with these placeholders in Sections~\ref{sec:prompt_source} and~\ref{sec:prompt_esbmc}. For readability purposes, we remove newline characters \texttt{\textbackslash{n}} and the backtick markers \texttt{```} that surround the source code and verifier output.

\begin{tcolorbox}[colback=green!5!white,colframe=green!75!black,title=Listing 1: Old Prompt Template used in ESBMC-AI]
    You are a secure code generator that parses vulnerable source code and output from a program called ESBMC, which contains vulnerability information about the source code. You should use the output from ESBMC to find the problem and correct the source code. ESBMC is always correct. You shall add a NULL check for every heap allocation you make. From this point on, you can only reply using the source code. You shall only output source code as a whole. The following text is the program's source code: \{source\} The following text is the output of ESBMC, reply OK if you understand: \{esbmc\} Generate a correction for the source code provided. Show the code only. Do not reply with acknowledgments.
\end{tcolorbox}

\subsubsection{Simple Prompts}
\label{sec:prompt_simple}

The prompt template in Section~\ref{sec:prompt_old} is quite long and forces a specific order in the information provided: first, the code, then the verifier feedback. For comparison reasons, we deem it necessary to consider shorter prompts and vary the code's and bug trace's relative position. More specifically, we include three pairs of prompts of different lengths, which contain the following information:
\begin{itemize}
    \item No verifier output provided;
    \item The verifier output is provided \textit{after} the source code;
    \item The verifier output is provided \textit{before} the source code.
\end{itemize}

For reference, we list all six prompt templates below.

\begin{tcolorbox}[colback=green!5!white,colframe=green!75!black,title=Listing 2: Simple Prompt Template with no ESBMC Output]
    The following source code segment might contain a memory vulnerability \{source\} Fix the source code segment.
    \tcblower
    Fix the memory vulnerability that may exist in the source code segment: \{source\}
\end{tcolorbox}

\begin{tcolorbox}[colback=green!5!white,colframe=green!75!black,title=Listing 3: Simple Prompt Template with ESBMC Output after Source Code]
    The following source code contains a memory vulnerability \{source\} The following is the output of ESBMC describing the vulnerability \{esbmc\}. Fix the source code.
    \tcblower
    Fix the source code: \{source\} \{esbmc\}
\end{tcolorbox}

\begin{tcolorbox}[colback=green!5!white,colframe=green!75!black,title=Listing 4: Simple Prompt Template with ESBMC Output before Source Code]
    ESBMC output describes a memory vulnerability in the source code; the following is ESBMC output: \{esbmc\} The following is the vulnerable source code: \{source\} Fix the source code.
    \tcblower
    Fix the source code: \{esbmc\} \{source\}
\end{tcolorbox}

\subsubsection{Persona Prompts}
\label{sec:prompt_persona}

Recent research has found that large language models produce better output when performing according to a specific role~\cite{white2023prompt}. This prompt engineering technique is usually called \textit{persona} and creates many optimization opportunities. In our case, we ask the following research question:

\begin{itemize}
    \item \textit{Does the role we assign to the large language model make a difference}.
\end{itemize}

Here are the six roles we compare:

\begin{enumerate}
    \item Programmer with 1 million years of experience;
    \item Senior software engineer;
    \item Automated code repair tool;
    \item Artificial intelligence that specializes in repairing C programs;
    \item The smartest human in the universe;
    \item Dog.
\end{enumerate}

Note that roles 1 and 2 are humanoid expert roles. Roles 3 and 4 are robotic expert roles. Roles 5 and 6 are wildcards, which were added as a control. These roles are inserted in the following prompt templates instead of the placeholder \texttt{\{role\}}.

\begin{tcolorbox}[colback=green!5!white,colframe=green!75!black,title=Listing 5: Persona Prompt Template no ESBMC Output]
    You’re a \{role\}. You’ll be shown some C code. Repair the code and display it. The code is \{source\}
    \tcblower
    From now on, act as an \{role\} that repairs AI C code. You will be shown AI C code. Provide the repaired C code as output, as would an \{role\}. Aside from the corrected source code, do not output any other text. The code is \{source\}
\end{tcolorbox}

\begin{tcolorbox}[colback=green!5!white,colframe=green!75!black,title=Listing 6: Persona Prompt Template with ESBMC Output after Source Code]
    You’re a \{role\}. You’ll be shown some C code, along with ESBMC output. Repair the code and display it. The code is \{source\} The ESBMC output is \{esbmc\}
    \tcblower
    From now on, act as an \{role\} that repairs AI C code. You will be shown AI C code, along with ESBMC output. Pay close attention to the ESBMC output, which contains a stack trace along with the type of error that occurred and its location. Provide the repaired C code as output, as would an \{role\}. Aside from the corrected source code, do not output any other text. The code is \{source\} The ESBMC output is \{esbmc\}
\end{tcolorbox}

\begin{tcolorbox}[colback=green!5!white,colframe=green!75!black,title=Listing 7: Persona Prompt Template with ESBMC Output before Source Code]
    You’re a \{role\}. You’ll be shown some C code, along with ESBMC output. Repair the code and display it. The ESBMC output is \{esbmc\} The source code is \{source\}
    \tcblower
    From now on, act as an \{role\} that repairs AI C code. You will be shown AI C code, along with ESBMC output. Pay close attention to the ESBMC output, which contains a stack trace along with the type of error that occurred and its location. Provide the repaired C code as output, as would an \{role\}. Aside from the corrected source code, do not output any other text. The ESBMC output is \{esbmc\} The source code is \{source\}
\end{tcolorbox}

\subsubsection{Source Code}
\label{sec:prompt_source}

Large language models can only process a limited input text length. In the case of ChatGPT~\cite{radford2019language}, which we use in our experiment of Section~\ref{sec:prompt_comparison}, the maximum input length is $16$K tokens. As such, it is often impossible to feed the large language model the whole code to be repaired. To overcome this limitation, we employ the following two strategies.

\paragraph{Contextual.} Most software verifiers, including ESBMC, do not just report the presence of a vulnerability, but they also include the line of code where it triggered an unwanted behavior. We cut the largest code window around the reported vulnerability to fit as much code as possible in the available space. More specifically, we select a window that contains around $90$\% lines of code before the vulnerability and $10$\% after.

\paragraph{One line.} For memory safety vulnerabilities, such as array-out-of-bounds, it is likely that modifying the very same line of code that triggered the verifier is sufficient to repair it. For this reason, we run a full set of experiments showing the large language model with only one line of code.

\subsubsection{Verifier Feedback}
\label{sec:prompt_esbmc}

The output of most software verifiers contains a bug report with the violated property and a full bug trace with a concrete value for each state (counterexample). Our question is whether these pieces of information are useful to the large language model. For this reason, we run two sets of experiments that include either the full counterexample or just the short bug report. We observe no significant difference between these.

\subsubsection{Prompt Combinations}

In summary, we have the following combinations of prompts:
\begin{itemize}
    \item \textbf{Without Verifier Feedback.} (2 simple + 2 persona * 6 role) * 2 source
    \item \textbf{With Verifier Feedback.} (1 old + 4 simple + 4 persona * 6 role) * 2 feedback * 2 source
\end{itemize}
for a total of $(2 + 2 * 6) * 2 + (1 + 4 + 4 * 6) * 2 * 2 = 144$ prompts.

\subsection{Comparing Templates}
\label{sec:prompt_comparison}

\begin{figure}[p]
\centering
    \begin{subfigure}[t]{0.68\textwidth}
    \centering
        \includegraphics[width=\textwidth]{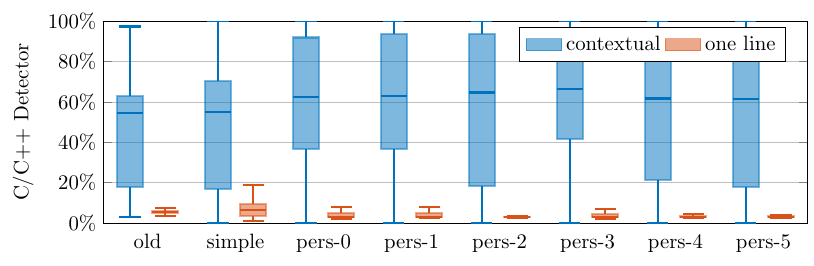}
        \caption{Prompt comparison.}
        \label{fig:lang_prompt}
    \end{subfigure}
    \hfill
    \begin{subfigure}[t]{0.31\textwidth}
    \centering
        \includegraphics[width=\textwidth]{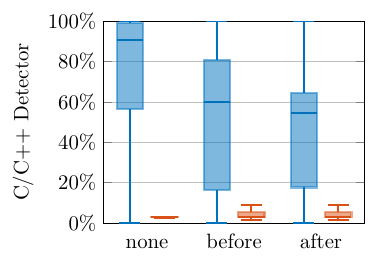}
        \caption{Verifier feedback.}
        \label{fig:lang_esbmc}
    \end{subfigure}
\caption{Distribution of probability scores from the C/C++ detector used in Visual Studio on the LLM repair patches. Persona prompts cause the LLM to produce C/C++ code more consistently.}
\label{fig:lang_results}
\end{figure}

\begin{figure}[p]
\centering
    \begin{subfigure}[t]{0.68\textwidth}
    \centering
        \includegraphics[width=\textwidth]{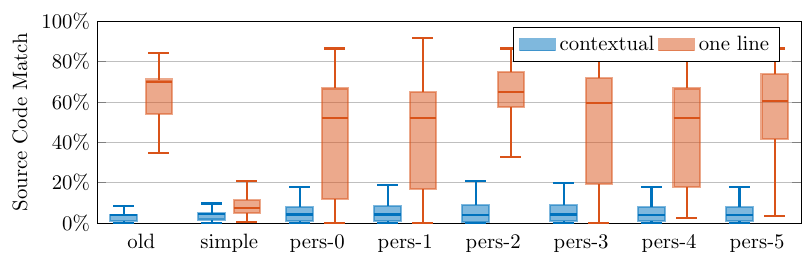}
        \caption{Prompt comparison.}
        \label{fig:match_prompt}
    \end{subfigure}
    \hfill
    \begin{subfigure}[t]{0.31\textwidth}
    \centering
        \includegraphics[width=\textwidth]{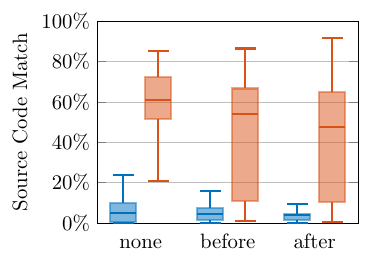}
        \caption{Verifier feedback.}
        \label{fig:match_esbmc}
    \end{subfigure}
\caption{String match between source code in the prompt and LLM output patch (ignoring whitespaces). When using the contextual strategy, the LLM tends to omit large swathes of the original code.}
\label{fig:match_results}
\end{figure}

\begin{table}[p]
\centering
\resizebox{\textwidth}{!}{%
\begin{tabular}{ |c|cccccccc|ccc| } 
    \hline
    & old & simple & pers-0 & pers-1 & pers-2 & pers-3 & pers-4 & pers-5 & none & before & after \\
    \hline
    contextual & 1.50\% & 1.25\% & 0.25\% & 0\% & 1.58\% & 2.16\% & 1.16\% & 1.25\% & 0.14\% & 1.32\% & 1.82\% \\
    \hline
    one line & 63.0\% & 16.91\% & 61.25\% & 61.41\% & 31.8\% & 56.3\% & 38.9\% & 52.0\% & 54.9\% & 41.7\% & 40.0\% \\
    \hline
\end{tabular}}
\caption{Percentage of repair patches that make the code compile. Asking the LLM to repair only one line of code yields patches that compile more consistently.}
\label{tab:compile_results}
\end{table}

\begin{table}[p]
\centering
\resizebox{\textwidth}{!}{%
\begin{tabular}{ |c|cccccccc|ccc| } 
    \hline
    & old & simple & pers-0 & pers-1 & pers-2 & pers-3 & pers-4 & pers-5 & none & before & after \\
    \hline
    contextual & 1.50\% & 1.25\% & 0.25\% & 0\% & 1.58\% & 2.16\% & 1.16\% & 1.25\% & 0.14\% & 1.32\% & 1.82\% \\
    \hline
    one line & 0.00\% & 4.25\% & 5.50\% & 5.83\% & 6.75\% & 8.25\% & 10.3\% & 10.5\% & 5.32\% & 9.57\% & 7.10\% \\
    \hline
\end{tabular}}
\caption{Percentage of repair patches that are successfully verified. All the patches that compile (see Table~\ref{tab:compile_results}) are also successfully verified when using the contextual strategy.}
\label{tab:verif_results}
\end{table}

This section compares the repair performance of LLMs using the different prompting strategies listed above. Overall, we measure the LLM performance on four metrics of increasing difficulty:
\begin{itemize}
    \item \textbf{Syntax.} Do the repair patches contain C code?
    \item \textbf{Relevance.} Do the repair patches match the input source code?
    \item \textbf{Compilation.} Do the repair patches compile?
    \item \textbf{Verification.} Do the repair patches solve the memory safety vulnerability?
\end{itemize}

\paragraph{Experimental Setup} The experiments were run on a distributed computational infrastructure. For the duration of the experiments, GPT-3.5-Turbo was used as the LLM of choice, specifically \texttt{gpt-3.5-turbo-0125}, with a temperature of $1.0$, which is the default set by the owners of the LLM API; no defaults were changed. ESBMC v7.4.0 was run on a server with Intel(R) Xeon(R) CPU E5-2620 v4 @ 2.10GHz, 198 GB RAM. For each of the $144$ prompt combinations listed in Section~\ref{sec:prompt_engineering}, we run $100$ random samples from the AI code dataset we generate in Section~\ref{sec:expanded_dataset}, for a total of $14400$ runs.

\paragraph{Syntax of the LLM Patches} As a first sanity check, we want to confirm that the LLM produces C code rather than a mixture of textual instructions and small code snippets. We do so by running the automated code detector used in Visual Studio\footnote{https://github.com/yoeo/guesslang} and extracting the score associated with \texttt{C} or \texttt{C++}. The results in Figure~\ref{fig:lang_results} illustrate markedly higher scores for a prompt containing a contextual window of source code rather than a single line. This is expected as the detector struggles to detect the language of short code snippets. Still, we can see that persona prompts yield a higher percentage of C-like LLM outputs for contextual strategy than other prompts (see Figure~\ref{fig:lang_prompt}. Also, adding the ESBMC output to the prompt makes things worse for contextual strategy but better for single lines of code (Figure~\ref{fig:lang_esbmc}).

\begin{tcolorbox}
\textbf{Syntax Takeaway.} Large language models produce C-like code fairly consistently.
\end{tcolorbox}

\paragraph{Relevance of the LLM Patches} We know that the vulnerabilities introduced by Mull \cite{8411727} are due to a few character changes only, e.g., replacing \texttt{<} with \texttt{<=}. Thus, we expect a successful patch to copy most of the input code verbatim, except for a small difference. To see whether the LLM repair matches our expectations, we measure how many characters the input and output code have in common (not counting whitespaces) and report the results in Figure~\ref{fig:match_results}. The contextual strategy has a low match, as the LLM sometimes reports only a few lines of code around the bug rather than copying the whole input code. One-liners (single) have a better match as it is easier for the LLM to repeat what we feed as input. The anomaly for single and simple prompts in Figure~\ref{fig:match_prompt} may be due to temporary changes in the ChatGPT backend that are out of our control rather than a real effect of our prompting technique \cite{chen2023chatgpt}. Again, adding feedback from ESBMC to our prompt seems to lower the performance of the LLM (see Figure~\ref{fig:match_esbmc}).

\begin{tcolorbox}
\textbf{Relevance Takeaway.} Large language models skip over crucial portions of the input code.
\end{tcolorbox}

\paragraph{Compiling the Repaired Code} Even though we explicitly ask the large language model to produce valid C code as its output (see Section~\ref{sec:prompt_engineering}), there is no mechanism to guarantee that it does so. For this reason, we check whether introducing the repaired patch back into the original code yields a piece of code that compiles. The results in Table~\ref{tab:compile_results} show that most of the runs did not produce compile code. However, there is a marked difference between how much source code is shown to the large language model. Indeed, letting the large language model repair only one line of code yields around $50$\% compilable patches across all settings.

\begin{tcolorbox}
\textbf{Compilation Takeaway.} Multi-line patches hardly compile, one-liners have a $~50\%$ chance.
\end{tcolorbox}

\paragraph{Verifying the Repaired Code} Finally, we evaluate whether the repaired code passes all checks for memory safety vulnerabilities. We do so by invoking ESBMC on the repaired code and counting how many programs are successfully verified (see Table~\ref{tab:verif_results}). All the repair patches generated using the contextual strategy display an interesting property: if the code compiles, it is also successfully verified. However, while one-line repair patches might cause a verification failure or a time-out, they yield more successful repairs overall. In this respect, persona prompt techniques (no matter the specific role chosen) are the best.

\begin{tcolorbox}
\textbf{Verification Takeaway.} The best setting achieves less than 10\% successful repairs.
\end{tcolorbox}

\subsection{Comparing Individual Prompts}
\label{sec:best_prompt_comparison}

The following section will compare all the individual prompts executed in the experiments and outline the best prompt for the Contextual and Single-line experiments. The grouping of the experiments and the use of  ``prompt'' in this section refers to substituting the role and the ESBMC output type into a prompt template. The notation of a prompt is $x.y.z$ where $x$ describes the prompt index, $y$ describes the role index used\footnote{Incidentally, non Persona prompts are always index 0 for role.}, and $z$ describes the ESBMC output type. 

\begin{figure}[t]
\centering
    \begin{subfigure}[t]{0.45\textwidth}
        \centering
        \includegraphics[width=\textwidth]{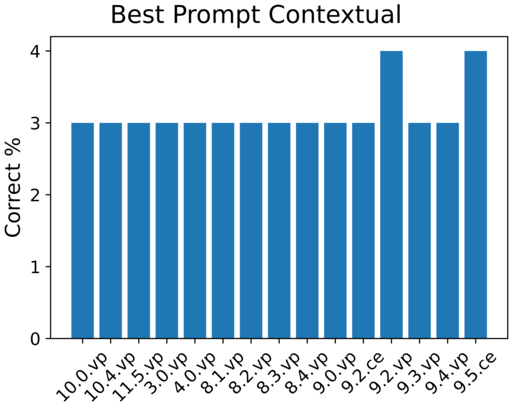}
        \caption{Contextual experiments.}
        \label{fig:best_prompt_baseline}
    \end{subfigure}
    \hfill
    \begin{subfigure}[t]{0.45\textwidth}
        \centering
        \includegraphics[width=\textwidth]{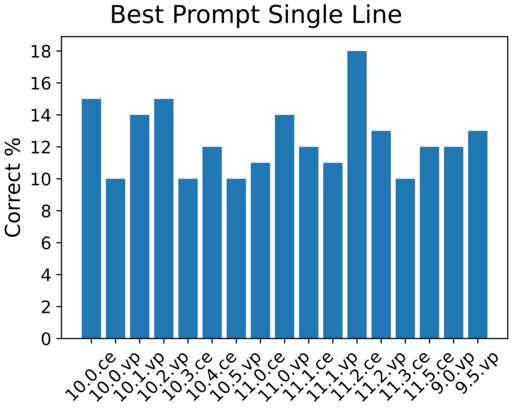}
        \caption{Single Line experiments.}
        \label{fig:best_prompt_single}
    \end{subfigure}
    \caption{Prompt comparison for the best prompts. The Contextual Figure presents results greater than or equal to 3. The Single Line Figure presents results greater than or equal to 10.}
\end{figure}

\paragraph{Contextual} Figure \ref{fig:best_prompt_baseline} shows the performance of the best prompts for the Contextual experiments. As evaluated in Section~\ref{sec:prompt_comparison}, it was discovered that Persona prompts performed overall better than the simple prompts; however, surprisingly, the Persona prompt assigned roles did not seem to affect the performance in an expected way. One note is that many prompts in the Contextual experiments did not successfully repair even one sample, so they have been omitted and not shown in Figure~\ref{fig:best_prompt_baseline}. Two prompts perform the best, they both use prompt template 9. The roles are ``Automated Code Repair Tool'' and ``Dog''. The verifier output can both be violated property (VP) or counterexample (CE), as they both successfully repaired 4\% of the samples.



\paragraph{Single Line} Regarding the best prompt in~\cref{fig:best_prompt_single}, the flipped Persona prompt template, index 11, was the most successful, at around $18\%$. The role that was used by the prompt was ``Automated Code Repair Tool''. The ESBMC output type used in the prompt was Counterexample (CE). The implications of these results suggest that, while assigning a specific role to the LLM is important, the role assigned may not impact performance consistently. However, it is worth noting that the ``Automated Code Repair Tool'' role seemed to have been among the best prompts in both Contextual and Single Line experiments. Also, longer prompts seemed to perform better in both Contextual and Single Line experiments.

\begin{tcolorbox}
\textbf{Best Prompt Takeaway.} The best prompt is the second prompt in Listing 7 in \cref{sec:prompt_persona}. Using the role ''Automated Code Repair Tool``.
\end{tcolorbox}

\subsection{Further Fine-Tuning of the Best Prompts}
\label{sec:best_prompts_finetuning}




Prompts can be further refined to achieve better results. Here, we focus on the best two prompts found in \cref{sec:best_prompt_comparison}: 9 and 11. Specifically, we consider two potential improvements. First, in Appendix~\ref{sec:backticks}, we explore the role of backticks, which are usually added as a separator around code snippets. Second, in the current Section, we present two prompts with longer and more detailed instructions, which in previous experiments have yielded better performance (see \cref{sec:llm_prompting}).

More specifically, we add further instruction to prompt templates 9 and 11 to do an additional 3 things. Firstly, we explicitly state to use the verifier output to find out what the fault of the sample is. Secondly, we instruct the LLM to repair only one line of code. Lastly, we instruct the LLM to minimize the size of the changes. 

The modified prompt templates, referred to as 9-2 and 11-2 can be seen below:

\begin{tcolorbox}[colback=green!5!white,colframe=green!75!black,title=Prompt Template 9-2 and Prompt Template 11-2]
    From now on, act as an \{role\} that repairs AI C code. You will be shown a line of AI C code, along with ESBMC output. Pay close attention to the ESBMC output, which contains what type of error has occurred and its location. Provide the repaired C code as output, as would an \{role\}. Aside from the corrected line of source code, do not output any other text. The code is \{source\_code\} The ESBMC output is \{esbmc\_output\} Guideline: Always prefer to repair using a single line of C code, unless necessary. Guideline: Read the error in the ESBMC output and try to repair the fault.
    \tcblower
    From now on, act as an \{role\} that repairs AI C code. You will be shown a line of AI C code, along with ESBMC output. Pay close attention to the ESBMC output, which contains what type of error has occurred and its location. Provide the repaired C code as output, as would an \{role\}. Aside from the corrected line of source code, do not output any other text. The ESBMC output is \{esbmc\_output\} The source code is \{source\_code\} Guideline: Always prefer to repair using a single line of C code, unless necessary. Guideline: Read the error in the ESBMC output and try to repair the fault.
\end{tcolorbox}

The impact of the new prompts will be presented in the following Section.

\subsection{Lessons Learned}
\label{sec:prompt_lessons}

As shown in \cref{sec:prompt_comparison}, repairing out-of-distribution code with large language models appears to be a delicate endeavour. On the one hand, some advanced prompt engineering techniques, e.g., assigning the model a persona role, seem to improve the success rate overall. However, the specific role does not seem to matter: the performance of ``Dog'' is more or less the same as that of a more credible role such as ``Automated code repair tool''.

On the other hand, the choice of which information we show to the model seems crucial for obtaining good-quality patches. Perhaps surprisingly, providing feedback from a verifier about the nature of the vulnerability makes the performance worse. At the same time, choosing how many lines of code we include in the prompt is crucial for generating patches that compile and remove the vulnerabilities. Conversely, a cleverer choice of which lines of code to include in the prompt may be one of the most promising avenues for improvement. In this respect, formal methods such as static analysis can greatly help in identifying such program subsets.


Finally, the overall percentage of successfully repaired programs is just below $18$\%, even under the best settings. This suggests that it would take a large language model several attempts to propose a correct patch, thus greatly increasing the computational cost of automated program repair. We plan to minimize the cost of such an iterative approach in Work Package 3.

\section{Iterative Automated Program Repair}
\label{sec:esbmc_ai}



\begin{figure}[H]
    \centering
    \includegraphics[width=0.75\textwidth]{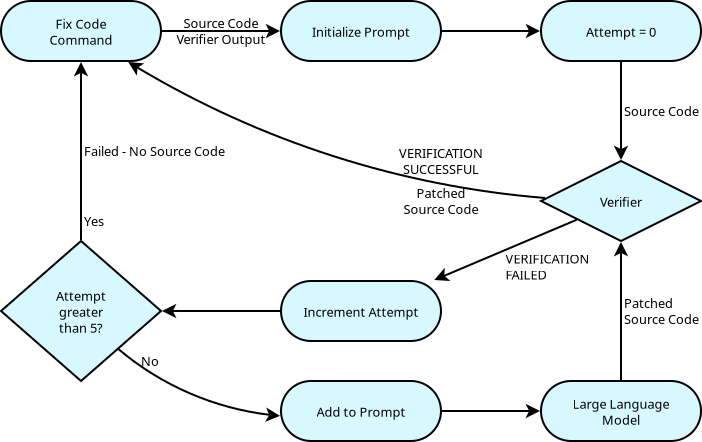}
    \caption{Diagram of the iterative repair algorithm of ESBMC-AI.}
    \label{fig:esbmc_ai_fcm}
\end{figure}

Allowing the LLM to have multiple attempts at repairing the faulty sample may yield an increase in automated code repair (APR) performance. \cref{fig:esbmc_ai_fcm} shows a diagram of how the iterative repair would function. It is an extended version of \cref{fig:singleattemptrepairdiagram} that adds an iterative loop element with the tracking of attempts. If the total attempts exceed the limit, in this case 5, then the verification fails.

With that in mind, we propose the following research questions:

\begin{enumerate}
    \item \textit{Do large language models combined with formal verification increase automated program repair performance of AI C code when given multiple attempts to do so?}
    \item \textit{Does showing the history of patches to the LLM improve the performance of iterative APR?}
    \item \textit{What is the optimal way to show the history to the LLM?}
    \item \textit{What is the optimal temperature to conduct APR of AI C Code?}
\end{enumerate}

Section~\ref{sec:esbmc_ai_methodology} will describe the implementation details of the experiments. Section~\ref{sec:esbmc_ai_setup} will describe the experimental setup used to conduct the experiments. Section~\ref{sec:esbmc_ai_eval} will analyze and interpret the results.

\subsection{Methodology and Workflow}
\label{sec:esbmc_ai_methodology}

By adding the iterative loop element to the experimental setup, we expect an increase in the repair performance since the LLM has multiple attempts to repair the artifact successfully. 

\paragraph{ESBMC-AI} The automated program repair functionality in ESBMC-AI builds a list of pairs of previously attempted repairs consisting of the source code and verifier output on top of the initial source code and verifier output \cite{charalambous2023new}. The iterative loop is described in Algorithm~\ref{alg:iterrepair} \cite{charalambous2023new}:

\begin{enumerate}
    \item The message list initially consists of the initial prompt and the source code, along with the verifier output, being substituted in. In the algorithm, Line~\ref{alg:iterrepair:1} initializes an empty message prompt, and on Line~\ref{alg:iterrepair:2} gets assigned to as described.
    \item The LLM is sent the prompt instructing it to repair the source code, using the verifier output as a guide, as seen in Line~\ref{alg:iterrepair:3}.
    \item The LLM returns the patched source code, and ESBMC is then tasked with verifying it is free of memory violations. If it is correct, then the APR process has successfully concluded.
    \item If the verifier confirms that the patched code is still wrong, a new message is added to the message list consisting of a new prompt template with the latest source code, and new verifier output is also included.
    \item The process repeats (from step 2).
\end{enumerate}

\begin{algorithm}[t]
    \caption{Iterative APR}
    \label{euclid}
    \begin{algorithmic}[1] 
        \Procedure{FIX\_CODE}{$P,s,T,AI,V$} \Comment{Tries to repair $s$ in $T$ attempts or less using $AI$. The prompt template is represented by $P$}
            \State $m\gets []$\label{alg:iterrepair:1}
            \For{$c\gets 0,T$} \Comment{Check if we have exhausted allocated attempts}
                \State $v,o\gets V(s)$ \Comment{Verify if $s$ does \textbf{\textit{not}} have memory vulnerabilities. The verifier verdict $v$ and feedback $o$}
                \If{$e$}
                    \State \textbf{return} $o$
                \Else
                    \State $m\gets P(m, s, e)$ \label{alg:iterrepair:2} \Comment{Creates a prompt using $s$ and $e$, and previous messages $m$}
                    \State $s\gets AI(m)$ \label{alg:iterrepair:3} \Comment{Call the LLM and get patched source code}
                \EndIf
            \EndFor
            \State \textbf{return} \textbf{\textit{null}} \Comment{No solution is found, all attempts exhausted}
        \EndProcedure
    \end{algorithmic}
    \label{alg:iterrepair}
\end{algorithm}


\subsubsection{Prompt Settings}

\paragraph{Prompt Templates} Prompts 9 and 11 will be used for the iterative repair experiments, along with prompts 9-2 and 11-2 from \cref{sec:best_prompts_finetuning}. Additionally, the old ESBMC-AI prompt will also be used in the experiments discussed in Section~\ref{sec:prompt_old}; the aim of including such a prompt is to observe the improvement in performance that the new prompts provide in an iterative environment, as it allows for an additional baseline measurement.

\paragraph{Source Code} Due to the introduction of the iterative APR loop, the prompt structure becomes incompatible with the contextual experiments conducted in Section~\ref{sec:llm_prompting}. The reason for this is that the contextual experiments were conducted to maximize the amount of source code and verifier output placed into the prompt. By its very nature, the creation of a contextual source code system with iterative loop mechanics would not work, as each iteration would require the space that the initial prompt had already taken. For the experiments in this section, the one-line format of displaying the source code has been chosen, as described in Section~\ref{sec:prompt_source}. The one-line experiments were also the group of experiments that performed better than contextual in every metric, as seen in Section~\ref{sec:prompt_comparison}.

\paragraph{Verifier Output} In the previous experiments, the output of the verifier was either to show the counter-example stack trace and violated property, denoted as CE, or show only the violated property section of the verifier output, denoted as VP, as discussed in Section~\ref{sec:prompt_esbmc}. Due to the structure of the AI C Code being repaired, the stack-trace generated is very large. Thus, the context window of the LLM fills up without going through all the cycles of the iterative APR loop, making running the AI C code experiments infeasible. For the iterative experiments, the VP output type was chosen, as it discards the large stack-trace produced by the verifier, instead keeping the violated property that contains a copy of the statement that the error occurs in, along with the type of error that the verifier had detected during the verification process. 

\subsubsection{Message History}

\begin{figure}
    \centering
    \includegraphics[width=0.55\linewidth]{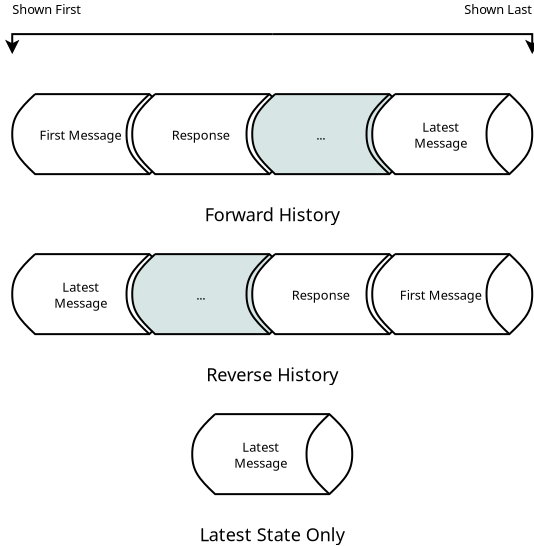}
    \caption{The 3 message history formats used in the experiments visualized.}
    \label{fig:esbmc_ai_message_history_formats}
\end{figure}

The iterative APR process introduces the concept of a message history: the collection of previous messages sent to the LLM. The three methods of representing current and previous messages are denoted as \textit{Latest State Only}, \textit{Forward History}, and \textit{Reverse History}. The next paragraphs provide detailed explanations of each. \cref{fig:esbmc_ai_message_history_formats} visually represents each format.

\paragraph{Latest State Only Experiments (LSO)} The LSO experiments are conducted to establish a baseline performance for the iterative APR approach. The iterative repair cycle of ESBMC-AI is modified to discard the history at each process of the APR loop. Effectively making it so that any previous repairs, aside from the last patch and verifier output, are discarded and not visible to the LLM.

\paragraph{Forward History Experiments} The Forward History experiments construct the prompt sent to the LLM such that the messages are in chronological order. In addition, the experiments store information from previous messages. The LLM gets a complete history of prior messages from the start of the repair process until the end.

\paragraph{Reverse History Experiments} Much like the Normal History experiments described previously, the Reverse History experiments use the same principle. However, the main difference is that the messages are reversed. The idea behind this is that the original state of the code would be displayed last in the message list. The last message read by the LLM would be the original state, which can potentially stop the LLM from drifting too far from the original code state.

\subsection{Experimental Setup}
\label{sec:esbmc_ai_setup}

This section will explore the results of the iterative code repair experiments.

\paragraph{Hardware} The iterative loop experiments will be conducted using the following environment:

\begin{itemize}
    \item The experiments were conducted on a server running both Intel Ivybridge and Haswell CPUs with 32 GB of RAM.
    \item ESBMC-AI version 0.5.0rc4
    \item ESBMC version 7.4.0 64-bit x86\_64 linux
    \item The LLM chosen is GPT-3.5-Turbo
    \item Each experiment was conducted over the 100 samples that the previous experiments used, in Section~\ref{sec:llm_prompting}.
\end{itemize}

Each experiment will be carried out over the following temperatures: \textit{0.0, 0.4, 0.7, 1.0, 1.3}. Each of the temperatures aims to determine which temperature constitutes the best performance for AI C code APR.

\subsection{Experimental Evaluation}
\label{sec:esbmc_ai_eval}

\subsubsection{Successful Verifications Per Attempt}

\begin{figure}[H]
    \centering
    \begin{subfigure}[t]{0.32\textwidth}
        \centering
        \includegraphics[width=\textwidth]{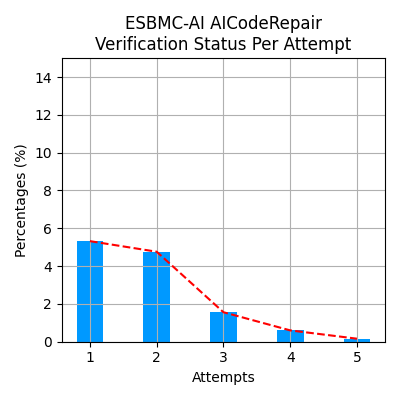}
        \caption{Latest State Only.}
    \end{subfigure}
    \hfill
    \begin{subfigure}[t]{0.32\textwidth}
        \centering
        \includegraphics[width=\textwidth]{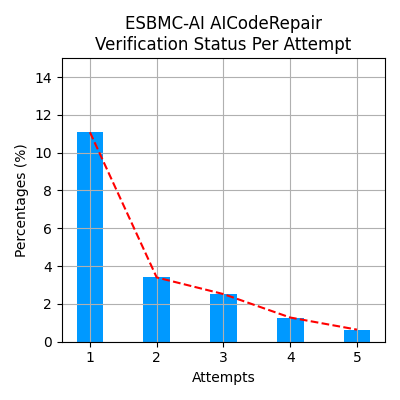}
        \caption{Forward History.}
    \end{subfigure}
    \hfill
    \begin{subfigure}[t]{0.32\textwidth}
        \centering
        \includegraphics[width=\textwidth]{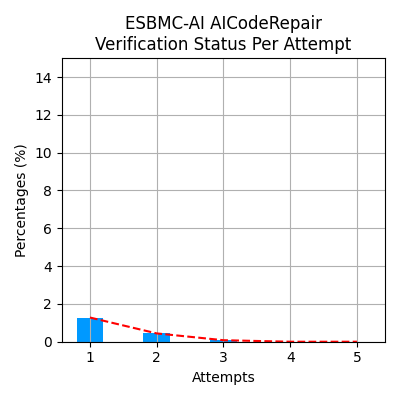}
        \caption{Reverse History.}
    \end{subfigure}
    \caption{Successful verifications by attempt.}
    \label{fig:esbmc_ai_successful_by_attempt}
\end{figure}

\cref{fig:esbmc_ai_successful_by_attempt} illustrates the percentage of successful repairs at each attempt after the first, the total being the overall amount of samples repaired. The Forward History experiment performed best, followed by the LSO experiment and the Reverse History. The Forward History experiments had the most successful attempt at repairing the AI C code on the first try; from the 1st retry and onwards, the number of successful code repairs is significantly lowered. LSO had a similar number of 1st and 2nd attempt successful repairs of samples, dropping sharply on the 3rd attempt. This could be caused by the LLM changing the state of the line where the fault occurs too much by the 2nd attempt. Thus making it extremely unlikely that a correct solution will be found. Reverse History had the lowest performance and had no successful repairs by the 2nd retry.

\begin{tcolorbox}
\textbf{Optimal Attempts Takeaway.} There is a sharp drop in successfully repairing a sample after the initial attempt.
\end{tcolorbox}

\subsubsection{Successful Repairs Per Temperature}

\begin{figure}[H]
    \centering
    \begin{subfigure}[t]{0.32\textwidth}
        \centering
        \includegraphics[width=\textwidth]{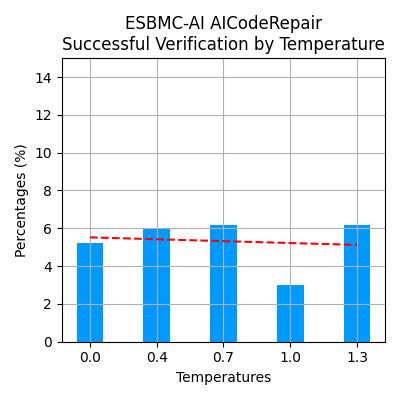}
        \caption{Latest State Only.}
    \end{subfigure}
    \hfill
    \begin{subfigure}[t]{0.32\textwidth}
        \centering
        \includegraphics[width=\textwidth]{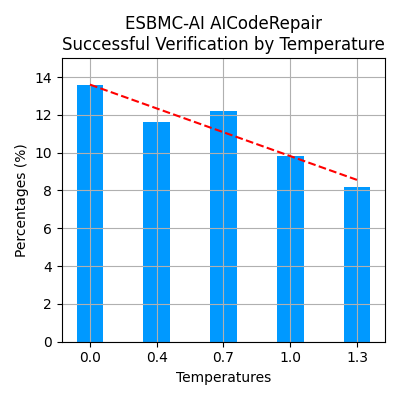}
        \caption{Forward History.}
    \end{subfigure}
    \hfill
    \begin{subfigure}[t]{0.32\textwidth}
        \centering
        \includegraphics[width=\textwidth]{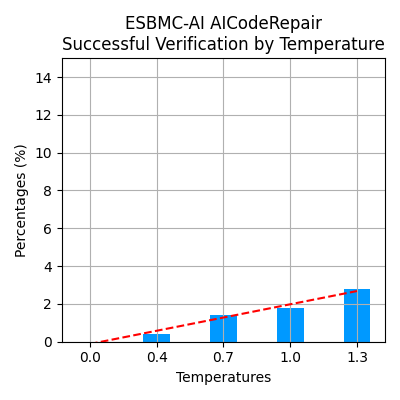}
        \caption{Reverse History.}
    \end{subfigure}
    \caption{Successful verifications by temperature.}
    \label{fig:esbmc_ai_successful_per_temp}
\end{figure}

Figure~\ref{fig:esbmc_ai_successful_per_temp} illustrates each experiment set's successful repairs per temperature. The main observation of each experiment is that temperature affects the number of successful repairs differently. Varying the temperature does not increase the repair performance for LSO. In contrast, the Forward History experiments have an inverse correlation between increasing the temperature and getting more repairs. In other words, the lower temperature values, which make the LLM behave more deterministically, yield higher repair performance. Interestingly, the opposite is true for Reverse History. While the performance of Reverse History is the lowest of the three experiments, there is a direct correlation between the number of successful repairs and a higher temperature. 

\begin{tcolorbox}
\textbf{Temperature Takeaway.} A lower temperature yields a higher repair accuracy for AI C Code APR using LSO or Forward History.
\end{tcolorbox}

\subsubsection{Successful Verifications Per Prompt}

\begin{figure}[H]
    \centering
    \begin{subfigure}[t]{0.32\textwidth}
        \centering
        \includegraphics[width=\textwidth]{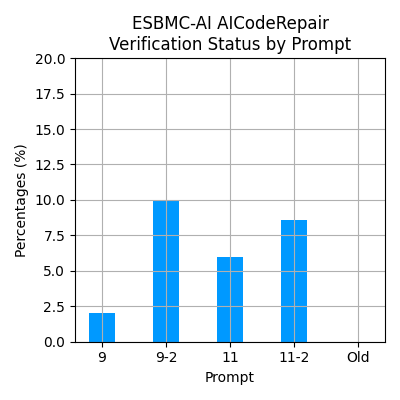}
        \caption{Latest State Only.}
    \end{subfigure}
    \hfill
    \begin{subfigure}[t]{0.32\textwidth}
        \centering
        \includegraphics[width=\textwidth]{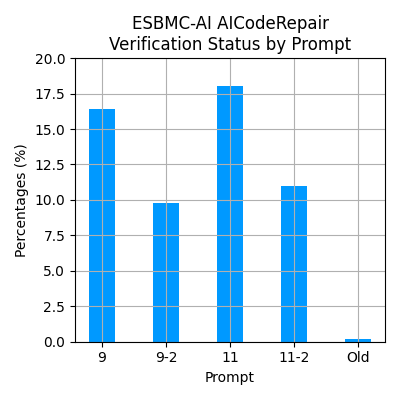}
        \caption{Forward History.}
    \end{subfigure}
    \hfill
    \begin{subfigure}[t]{0.32\textwidth}
        \centering
        \includegraphics[width=\textwidth]{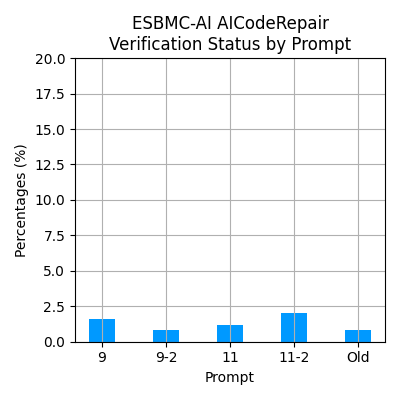}
        \caption{Reverse History.}
    \end{subfigure}
    \caption{Successful verifications by prompt.}
    \label{fig:esbmc_ai_successful_per_prompt}
\end{figure}

Figure~\ref{fig:esbmc_ai_successful_per_prompt} illustrates the percentage of successful repairs per prompt for the three experiments. These percentages represent the entire range of temperatures. The LSO experiments show prompts 11 and 11-2 performing the best, possibly due to the added context of the prompt. As the LSO experiments do not contain historical patches, they help repair performance by providing additional instruction. Interestingly, the Forward History experiment shows the opposite; prompts 9 and 11 perform best. Prompts 9 and 11 were the best in Section~\ref{sec:llm_prompting} experiments. The reverse history shows no significantly better performance between any of the prompts; in general, it performs the worst of all the experiments. In all three experiments, the Old prompt failed to repair many prompts successfully; however, the reverse one was the most successful.

Furthermore, we can observe the following if the new prompts are grouped into two types: modified and unmodified. The unmodified prompts have the highest amount of successful repairs in the Forward History experiments. However, the modified prompts perform better in the LSO experiments. In the end, the best prompts are, by far, prompt 9 and prompt 11 in the Forward History experiments. The most stable prompts are prompts 9-2 and 11-2, due to the added instructions, as they perform comparatively between the LSO and Forward History experiments.

\begin{tcolorbox}
\textbf{Best Prompts Takeaway.} The best prompts are prompt 9 and prompt 11.
\end{tcolorbox}

\subsection{Lessons Learned}
\label{sec:esbmc_ai_best}

Iterative APR allows the LLM to make multiple attempts to fix a source file. As seen from the experiments, the best results of non-iterative APR were $\sim$18\%, while the iterative approach explored in \cref{sec:esbmc_ai} increased the successful repairs to $\sim$25\%. Throughout all the experiments conducted, the option to produce more repaired samples is to use a lower temperature, with $0.0$ being the best for prompts 9 and 11. The best number of retries is 2, which means that the LLM has three attempts at resolving a single fault. The best type of history is the Forward History, where the LLM gains an advantage from past patches. LSO performed second best, but missing the history of repairs in the prompt leaves the LLM directionless in repairing the code.

In summary, we can answer our research questions as follows:

\begin{enumerate}
    \item \textit{Do large language models combined with formal verification increase automated program repair performance of AI C code when given multiple attempts to do so?} LLMs with formal verification increase the automated program repair performance when given multiple attempts. The best number of attempts is 3 for prompts 9 and 11, as a successful repair becomes unlikely after the 2nd retry.
    \item \textit{Does showing the history of patches to the LLM improve the performance of iterative APR?} Yes, when comparing the number of successful repairs between the LSO experiments and the Forward History experiments, the latter has a much higher number of successful repairs.
    \item \textit{What is the optimal way to show the history to the LLM?} The best format to show the history is to display the oldest messages first and the last message being the latest. This has been observed when comparing Forward History and Reverse History.
    \item \textit{What is the optimal temperature to conduct APR of AI C Code?} The optimal temperature to achieve the highest number of repaired samples is $0.0$ for prompts 9 and 11 for Forward History. A higher temperature is necessary for less conventional prompts to allow the LLM to parse it correctly.
\end{enumerate}

\section{Conclusions and Future Work}
\label{sec:conclusion}

In this report, we expanded NeuroCodeBench to create a large dataset of memory-vulnerable AI C code using mutations. We used ESBMC to classify which samples contained memory vulnerabilities and which samples were secure.

In addition, we used GPT-3.5-Turbo and various prompt engineering techniques to explore how well an LLM could repair the mutated code. In the process, we discovered that a long persona prompt with the role Automated Code Repair Tool is the most optimized at repairing AI C Code. Furthermore, we proposed methods for extracting the faulty source code from the large volume of AI C code to circumvent the issues that arise due to LLMs' relatively small context window.

Lastly, we used ESBMC-AI to test how the iterative APR process improves repair performance. We have showed that the iterative repair process provides a substantial increase of 7\% in repair performance. We also found that after attempt 3 the chances of successfully repairing a sample decrease significantly by 23\%.

In the future, we plan to conduct more experiments using a diverse set of LLMs to discover whether our findings generalize beyond GPT-3.5-Turbo. In this respect, open-source LLMs would benefit our research, as they can be fine-tuned for program repair.

\bibliographystyle{abbrv}
\bibliography{references}

\appendix

\section{Testing Backtick Performance Effectiveness}
\label{sec:backticks}


\newcommand{\threebackticks}{\textasciigrave\textasciigrave\textasciigrave}

The annotation of source code using Markdown code block syntax has always been an implicit feature in ChatGPT and LLMs \cite{openaidocs1}. It is widely accepted that code should be surrounded by {\threebackticks} to mark where it begins and ends. In Section~\ref{sec:llm_prompting}, the single iteration experiments provided an understanding of how effective LLMs are at repairing AI C code vulnerabilities. To understand the effectiveness of backticks in the source code provided, the experiments will be executed again, with the only difference being the exclusion of backticks in the prompt. The goal of these experiments is to influence future prompt engineering design. If the results of the prompts without backticks are better than with backticks, then this could mean excluding them would yield better APR accuracy in future experiments, such as those conducted in Section~\ref{sec:esbmc_ai}.

Each of the aforementioned prompts in Section~\ref{sec:prompt_engineering} has been modified to exclude the Markdown backticks. This is done in order to be able to directly compare the performance with the previous experiments and understand the true impact that backticks have on the performance of prompts. The results are going to be explored for the remainder of this section.

\begin{figure}[H]
\centering
    \begin{subfigure}[t]{0.68\textwidth}
    \centering
        \includegraphics[width=\textwidth]{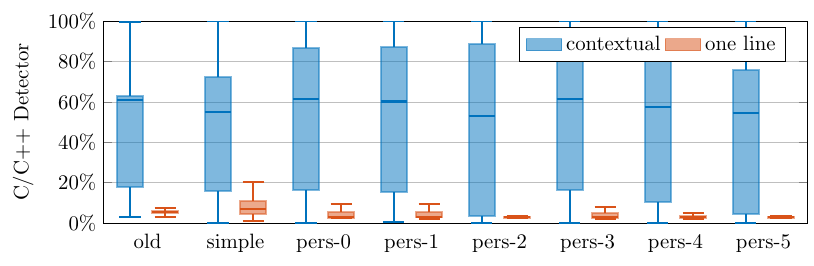}
        \caption{Prompt comparison.}
        \label{fig:btl_lang_prompt}
    \end{subfigure}
    \hfill
    \begin{subfigure}[t]{0.31\textwidth}
    \centering
        \includegraphics[width=\textwidth]{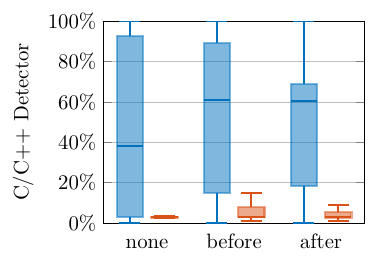}
        \caption{Verifier feedback.}
        \label{fig:btl_lang_esbmc}
    \end{subfigure}
\caption{Distribution of probability scores from the C/C++ detector used in Visual Studio on the LLM backtick-less repair patches. It exhibits a lower first quartile than the results with backticks included in the prompt, as illustrated in Figure~\ref{fig:lang_results}.}
\label{fig:btl_lang_results}
\end{figure}

\paragraph{C/C++ Detector Results} Figure~\ref{fig:btl_lang_results} illustrates the distribution of probability scores of the LLM output that the C/C++ detector assigned for each prompt, as described in Figure~\ref{fig:lang_results}. In each prompt shown in Figure~\ref{fig:btl_lang_prompt}, the main differences observed are that the experiments without backticks perform overall worse, aside from the simple prompts that did not display a noticeable difference. The old prompt contains a higher median value without backticks. And, the results for the persona prompts show a reduced confidence of the C/C++ detector, as observed by the Q1 and Q3 being lower, the median value was higher overall.

Figure~\ref{fig:btl_lang_esbmc} shows the detection results when the verifier is excluded from the prompt, along with the results for when the verifier is included before and after the source code. The syntax detector assigned the contextual experiments with no verifier output in the prompt less confidence in it being valid C/C++ source code, as can be observed by a lower Q1 and median score. When verifier output is included, the experiments with no backticks perform slightly better than the experiments with backticks. The results, for verifier output before the source code, exhibit a higher Q3. The results for verifier output after the source code exhibit a slightly higher median and Q3. The results for the contextual experiments signify the need to clearly mark the boundaries of code and raw output with Markdown syntax code blocks, as it shows that the LLM could output more consistent C/C++ code as evidenced by the detector confidence results.

The experiments that were conducted using one line source code prompts show no major difference in results for each of the prompts. The results for the verifier output show a slightly higher max value for the experiments that include verifier output. This can be explained as experimental noise, as the difference is not significant enough.

\begin{tcolorbox}
\textbf{Syntax Takeaway.} Including verifier output causes the LLM to output more accurate C/C++.
\end{tcolorbox}

\begin{figure}[H]
\centering
    \begin{subfigure}[t]{0.68\textwidth}
    \centering
        \includegraphics[width=\textwidth]{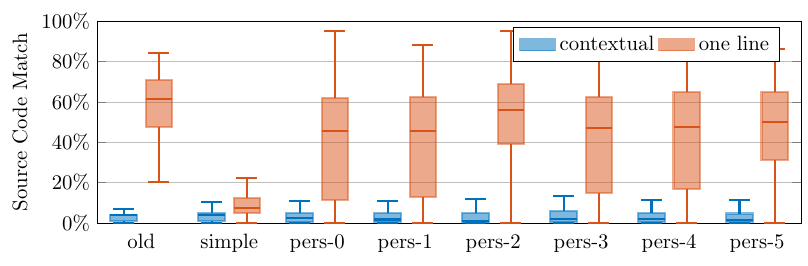}
        \caption{Prompt comparison.}
        \label{fig:btl_match_prompt}
    \end{subfigure}
    \hfill
    \begin{subfigure}[t]{0.31\textwidth}
    \centering
        \includegraphics[width=\textwidth]{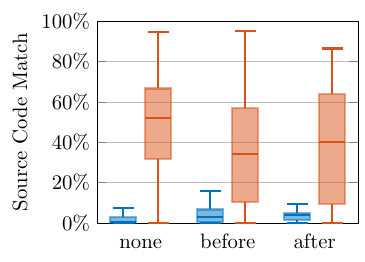}
        \caption{Verifier feedback.}
        \label{fig:btl_match_esbmc}
    \end{subfigure}
\caption{String match between source code in the prompt and LLM output patch (ignoring whitespaces). This figure is the equivalent of Figure \ref{fig:match_results} for the backtick-less experiment.}
\label{fig:btl_match_results}
\end{figure}

\paragraph{Relevance Results} The measurements that show how similar the proposed patches by the LLM were for the experiments without backticks can be observed in Figure \ref{fig:btl_match_results}. As stated in the experiments with backticks, it is expected that the LLM outputs the code verbatim, aside from the single line of code to be patched. Figure~\ref{fig:btl_match_prompt} illustrates a similarity box plot with the original code for each prompt. For the contextual experiments, the backtick-less match to the original prompt less in every prompt, as the max match values and Q3 values, and median values assigned in each prompt is lower for the backtick-less experiments. The results for one line mirror the results for contextual, aside from the fact that the overall difference is much more noticeable, and that some prompts have a higher max match value for backtick-less experiments.

Figure~\ref{fig:btl_match_esbmc} illustrates a box plot of how the results match the original for each type of verifier output experiment. As is observed, the performance seems lower in every metric. Contextual results have lower first\footnote{On plots where the first quartile is not zero.} quartiles, along with a lower median. For the verifier output experiments where no verifier output is included, there is a significantly less Q3 matching to the original prompt. For the one line experiments, experiments where no verifier output is included, and experiments where the verifier output is placed before source code, the max value of those experiments is higher for both, however, every other metric minimum value and Q1-3 are all lower.  

\begin{tcolorbox}
\textbf{Relevance Takeaway.} The LLM outputs code that more closely matches the original input when the backticks are included in the input.
\end{tcolorbox}

\paragraph{Compilation Results} The experiments conducted to find out the difference in the amount of samples that successfully compile, as the LLM produces valid C code as its output without Markdown backticks, is shown in Table \ref{tab:btl_compile_results}. The results for the contextual experiments are the exact same as with the Markdown backtick experiments, shown in Table \ref{tab:compile_results}. For the one line experiments, the experiments with backticks seem to perform worse, the prompts and the verifier output sections of the table both have a higher number of samples that successfully compile for the experiments that exclude the backticks from the prompts.

\begin{table}[h]
\centering
\resizebox{\textwidth}{!}{%
\begin{tabular}{ |c|cccccccc|ccc| } 
    \hline
    & old & simple & pers-0 & pers-1 & pers-2 & pers-3 & pers-4 & pers-5 & none & before & after \\
    \hline
    contextual & 1.50\% & 1.25\% & 0.25\% & 0\% & 1.58\% & 2.16\% & 1.16\% & 1.25\% & 0.14\% & 1.32\% & 1.82\% \\
    \hline
    one line & 42.5\% & 19.91\% & 62.7\% & 71.5\% & 36.7\% & 57.5\% & 32.25\% & 58.5\% & 59.9\% & 42.1\% & 43.3\% \\
    \hline
\end{tabular}}
\caption{Percentage of repair patches that make the code compile. Asking the LLM to repair only one line of code yields patches that compile more consistently. Much like with Markdown backticks.}
\label{tab:btl_compile_results}
\end{table}

\begin{tcolorbox}
\textbf{Compilation Takeaway.} With large samples of code, as seen in the contextual experiments, the LLM does not exhibit an increase in compilable code output, the amount is the exact same as with backticks. For smaller samples, compilable code is produced more often when backticks are excluded.
\end{tcolorbox}

\paragraph{Verifying the Code Results} Table \ref{tab:btl_verif_results} shows the verification success rate of each prompt and verifier feedback. For the contextual experiments, the results shown reflect the same pattern where the percentage of samples that successfully compiled is exactly the same as Tables \ref{tab:compile_results}, \ref{tab:verif_results}, and \ref{tab:btl_compile_results}. This implies that when the contextual experiments compile, they will also have a successful verification. For one line backtick-less experiments, the amount of samples that have been verified successfully is lower in all the metrics.

\begin{table}[h]
\centering
\resizebox{\textwidth}{!}{%
\begin{tabular}{ |c|cccccccc|ccc| } 
    \hline
    & old & simple & pers-0 & pers-1 & pers-2 & pers-3 & pers-4 & pers-5 & none & before & after \\
    \hline
    contextual & 1.50\% & 1.25\% & 0.25\% & 0\% & 1.58\% & 2.16\% & 1.16\% & 1.25\% & 0.14\% & 1.32\% & 1.82\% \\
    \hline
    one line & 0.00\% & 4.16\% & 5.00\% & 5.83\% & 4.25\% & 6.50\% & 9.25\% & 7.00\% & 5.21\% & 7.67\% & 5.10\% \\
    \hline
\end{tabular}}
\caption{Percentage of repair patches that are successfully verified. All the patches that compile (see Table~\ref{tab:compile_results}) are also successfully verified when using the contextual strategy.}
\label{tab:btl_verif_results}
\end{table}

\begin{tcolorbox}
\textbf{Verification Takeaway.} With large samples of code input, the code will compile and verify the same regardless of if code is surrounded in backticks or not. For smaller samples, the successful verification of code is slightly higher when backticks are included.
\end{tcolorbox}

\section{Iterative Repair Experiment Details}

\subsection{Successful Verifications By Prompt Per Temperature}

\begin{figure}[H]
    \centering
    \begin{subfigure}[t]{0.32\textwidth}
        \centering
        \includegraphics[width=\textwidth]{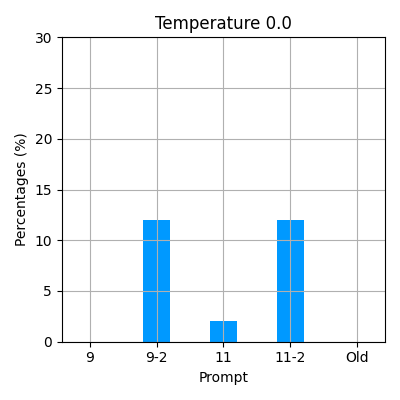}
        \caption{Latest State Only.}
    \end{subfigure}
    \begin{subfigure}[t]{0.32\textwidth}
        \centering
        \includegraphics[width=\textwidth]{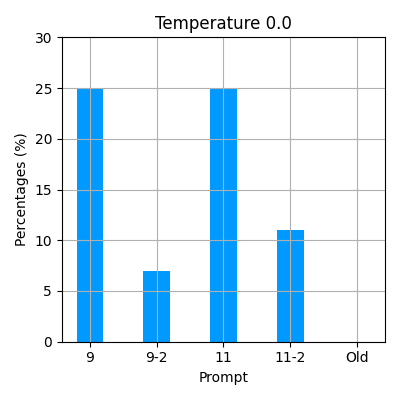}
        \caption{Forward History.}
    \end{subfigure}
    \hfill
    \caption{Successful verifications by prompt for temperature 0.0. Reverse History is not included as it did not yield any successful verifications.}
    \label{fig:esbmc_ai_successful_by_prompt_per_temp_0.0}
\end{figure}

\begin{figure}[H]
    \centering
    \begin{subfigure}[t]{0.32\textwidth}
        \centering
        \includegraphics[width=\textwidth]{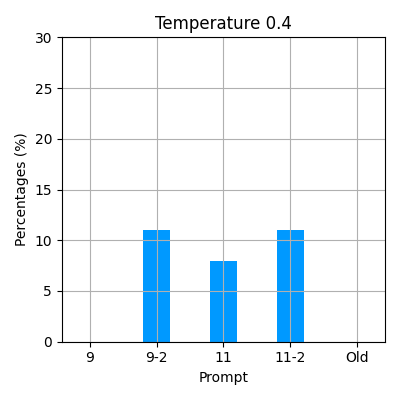}
        \caption{Latest State Only.}
    \end{subfigure}
    \hfill
    \begin{subfigure}[t]{0.32\textwidth}
        \centering
        \includegraphics[width=\textwidth]{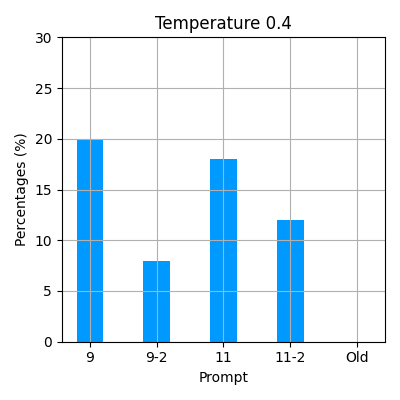}
        \caption{Forward History.}
    \end{subfigure}
    \hfill
    \begin{subfigure}[t]{0.32\textwidth}
        \centering
        \includegraphics[width=\textwidth]{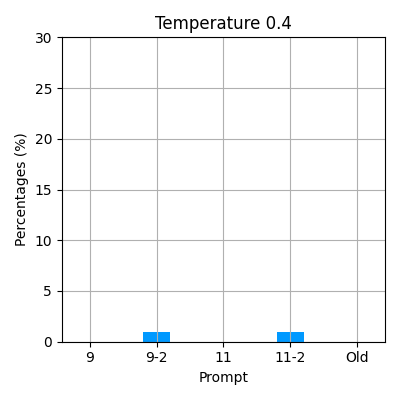}
        \caption{Reverse History.}
    \end{subfigure}
    \caption{Successful verifications by prompt for temperature 0.4.}
    \label{fig:esbmc_ai_successful_by_prompt_per_temp_0.4}
\end{figure}

\begin{figure}[H]
    \centering
    \begin{subfigure}[t]{0.32\textwidth}
        \centering
        \includegraphics[width=\textwidth]{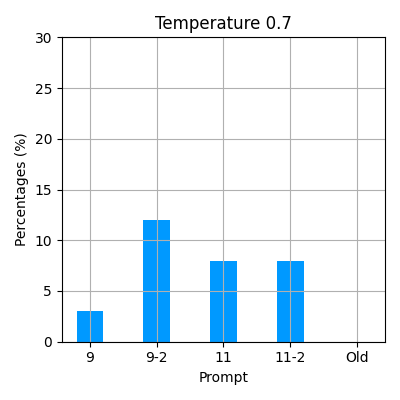}
        \caption{Latest State Only.}
    \end{subfigure}
    \hfill
    \begin{subfigure}[t]{0.32\textwidth}
        \centering
        \includegraphics[width=\textwidth]{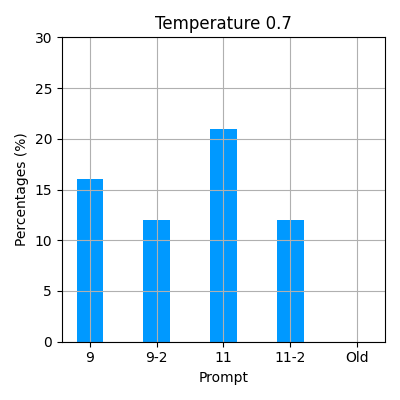}
        \caption{Forward History.}
    \end{subfigure}
    \hfill
    \begin{subfigure}[t]{0.32\textwidth}
        \centering
        \includegraphics[width=\textwidth]{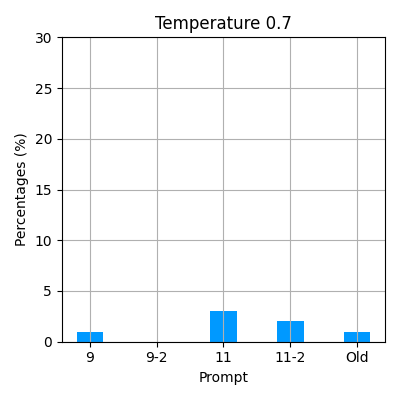}
        \caption{Reverse History.}
    \end{subfigure}
    \caption{Successful verifications by prompt for temperature 0.7.}
    \label{fig:esbmc_ai_successful_by_prompt_per_temp_0.7}
\end{figure}

\begin{figure}[H]
    \centering
    \begin{subfigure}[t]{0.32\textwidth}
        \centering
        \includegraphics[width=\textwidth]{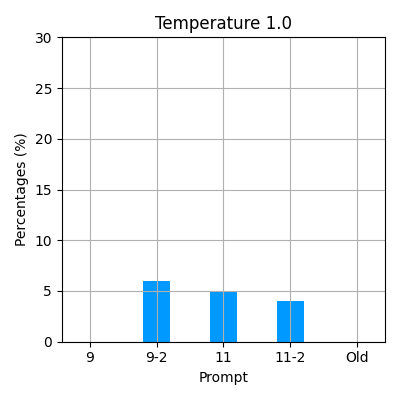}
        \caption{Latest State Only.}
    \end{subfigure}
    \hfill
    \begin{subfigure}[t]{0.32\textwidth}
        \centering
        \includegraphics[width=\textwidth]{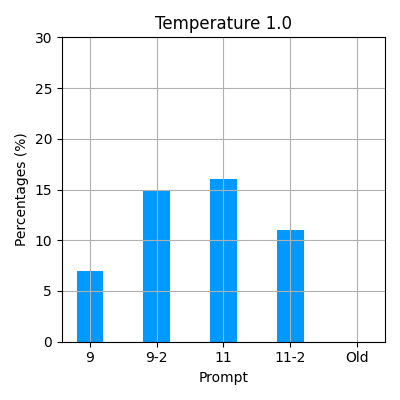}
        \caption{Forward History.}
    \end{subfigure}
    \hfill
    \begin{subfigure}[t]{0.32\textwidth}
        \centering
        \includegraphics[width=\textwidth]{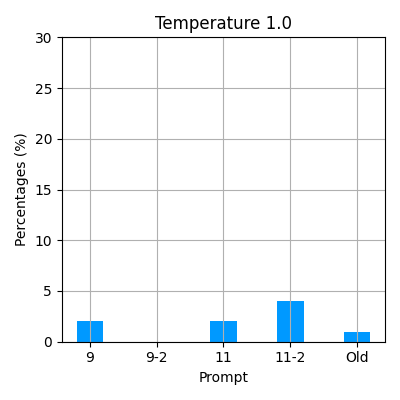}
        \caption{Reverse History.}
    \end{subfigure}
    \caption{Successful verifications by prompt for temperature 1.0.}
    \label{fig:esbmc_ai_successful_by_prompt_per_temp_1.0}
\end{figure}

\begin{figure}[H]
    \centering
    \begin{subfigure}[t]{0.32\textwidth}
        \centering
        \includegraphics[width=\textwidth]{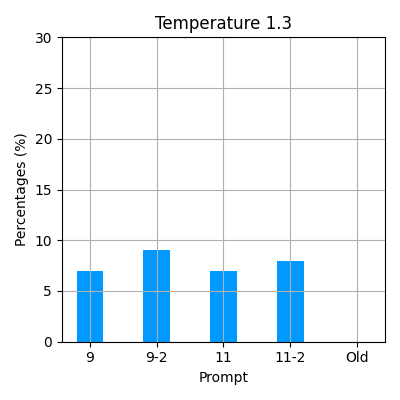}
        \caption{Latest State Only.}
    \end{subfigure}
    \hfill
    \begin{subfigure}[t]{0.32\textwidth}
        \centering
        \includegraphics[width=\textwidth]{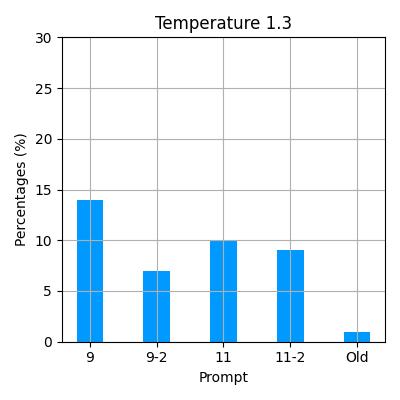}
        \caption{Forward History.}
    \end{subfigure}
    \hfill
    \begin{subfigure}[t]{0.32\textwidth}
        \centering
        \includegraphics[width=\textwidth]{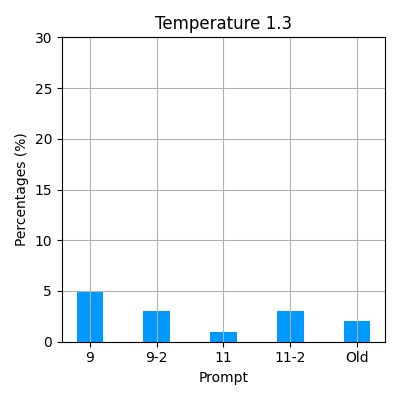}
        \caption{Reverse History.}
    \end{subfigure}
    \caption{Successful verifications by prompt for temperature 1.3.}
    \label{fig:esbmc_ai_successful_by_prompt_per_temp_1.3}
\end{figure}

The following figures are conducted over a range of temperatures, in order to further understand how the number of repairs are affected. \cref{fig:esbmc_ai_successful_by_prompt_per_temp_0.0,fig:esbmc_ai_successful_by_prompt_per_temp_0.4,fig:esbmc_ai_successful_by_prompt_per_temp_0.7,fig:esbmc_ai_successful_by_prompt_per_temp_1.0,fig:esbmc_ai_successful_by_prompt_per_temp_1.3} shows the successful verifications by prompt for temperatures 0.0, 0.4, 0.7, 1.0, and 1.3 respectively. The general pattern that can be observed, is that as the temperature is increased, the LSO and Forward History performance increases, while the opposite is true for Reverse History. For LSO, prompts 9-2, 11 and 11-2 perform the best, as temperature increases, so does the number of successful repairs. Prompts 9 and Old do not manage to repair any samples. The only exception is that in temperature 1.3, prompt 9 performed as well as prompt 11, however, still lower than prompts 9-2 and 11-2.

The best Forward History prompts are prompt 9 and prompt 11, they have the same number of successful repairs in temperature 0.0, which is the highest. Prompt 9 is only surpassed by the other prompts in \cref{fig:esbmc_ai_successful_by_prompt_per_temp_1.0,fig:esbmc_ai_successful_by_prompt_per_temp_0.7}. Aside from that, prompt 9 and 11 are the best performing prompts. Prompts 9 and 11 solve the exact same samples, however, the patches produced are different. 

As seen in the previous experiments, the Reverse History experiments score the lowest in successfully repairing AI C code. In \cref{fig:esbmc_ai_successful_by_prompt_per_temp_0.0}, Reverse History was unable to repair any samples. The reason for this is probably due to the indirect nature of displaying the history in reverse order, making the LLM perform worse as the instructions are not as clear for instruct models.

\begin{tcolorbox}
\textbf{Prompt Takeaway.} Forward History performs the best during iterative APR. Prompt 9 and 11 are the best prompts for APR, however, prompt 11 has the advantage of being more stable. It performs better than prompt 9 in LSO.
\end{tcolorbox}

\subsection{Successful Verifications By Attempt Per Temperature}

\begin{figure}[H]
    \centering
    \begin{subfigure}[t]{0.32\textwidth}
        \centering
        \includegraphics[width=\textwidth]{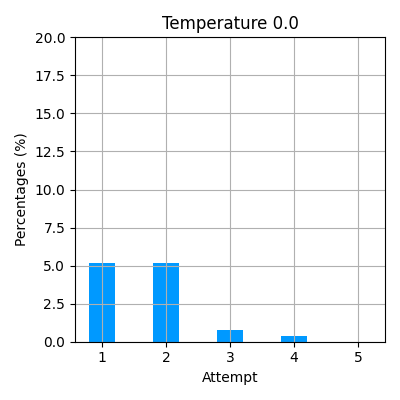}
        \caption{Latest State Only.}
    \end{subfigure}
    \begin{subfigure}[t]{0.32\textwidth}
        \centering
        \includegraphics[width=\textwidth]{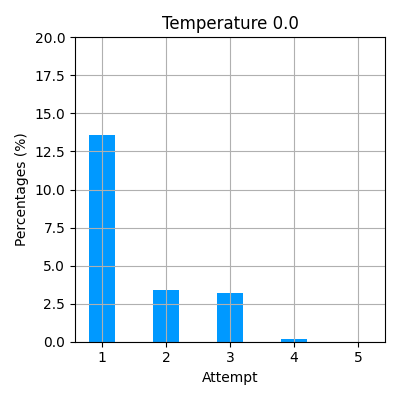}
        \caption{Forward History.}
    \end{subfigure}
    \hfill
    \caption{Successful verifications by attempt for temperature 0.0.}
    \label{fig:esbmc_ai_successful_by_attempt_per_temp_0.0}
\end{figure}

\begin{figure}[H]
    \centering
    \begin{subfigure}[t]{0.32\textwidth}
        \centering
        \includegraphics[width=\textwidth]{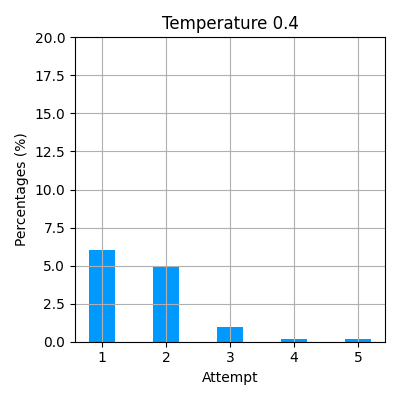}
        \caption{Latest State Only.}
    \end{subfigure}
    \hfill
    \begin{subfigure}[t]{0.32\textwidth}
        \centering
        \includegraphics[width=\textwidth]{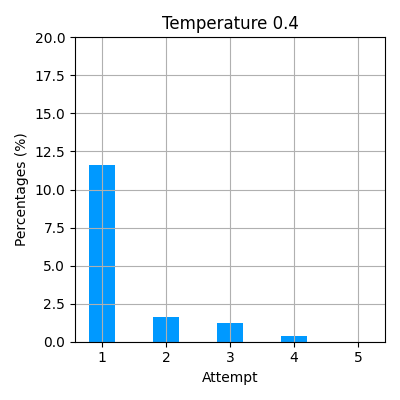}
        \caption{Forward History.}
    \end{subfigure}
    \hfill
    \begin{subfigure}[t]{0.32\textwidth}
        \centering
        \includegraphics[width=\textwidth]{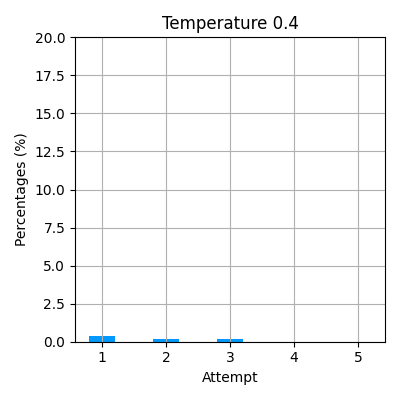}
        \caption{Reverse History.}
    \end{subfigure}
    \caption{Successful verifications by attempt for temperature 0.4.}
    \label{fig:esbmc_ai_successful_by_attempt_per_temp_0.4}
\end{figure}

\begin{figure}[H]
    \centering
    \begin{subfigure}[t]{0.32\textwidth}
        \centering
        \includegraphics[width=\textwidth]{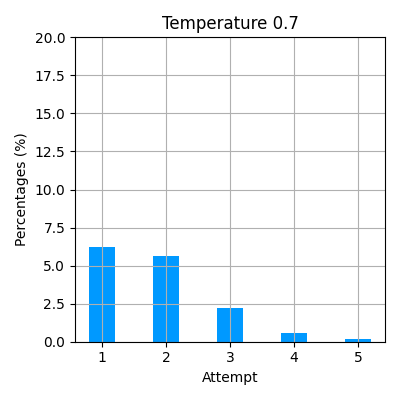}
        \caption{Latest State Only.}
    \end{subfigure}
    \hfill
    \begin{subfigure}[t]{0.32\textwidth}
        \centering
        \includegraphics[width=\textwidth]{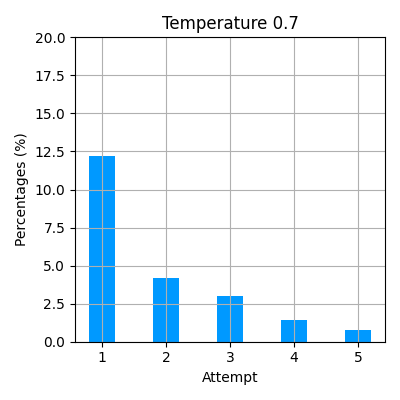}
        \caption{Forward History.}
    \end{subfigure}
    \hfill
    \begin{subfigure}[t]{0.32\textwidth}
        \centering
        \includegraphics[width=\textwidth]{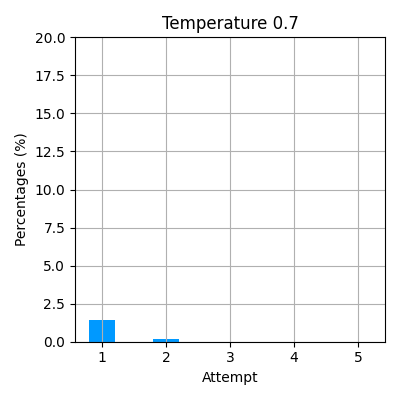}
        \caption{Reverse History.}
    \end{subfigure}
    \caption{Successful verifications by attempt for temperature 0.7.}
    \label{fig:esbmc_ai_successful_by_attempt_per_temp_0.7}
\end{figure}

\begin{figure}[H]
    \centering
    \begin{subfigure}[t]{0.32\textwidth}
        \centering
        \includegraphics[width=\textwidth]{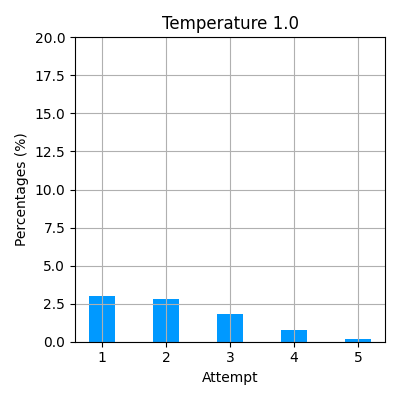}
        \caption{Latest State Only.}
    \end{subfigure}
    \hfill
    \begin{subfigure}[t]{0.32\textwidth}
        \centering
        \includegraphics[width=\textwidth]{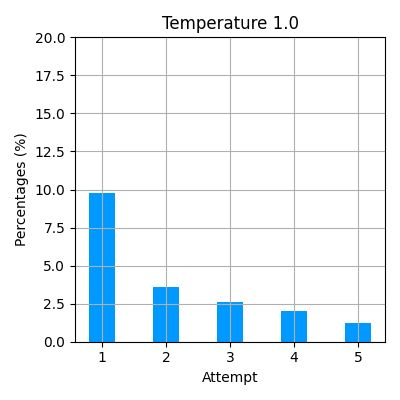}
        \caption{Forward History.}
    \end{subfigure}
    \hfill
    \begin{subfigure}[t]{0.32\textwidth}
        \centering
        \includegraphics[width=\textwidth]{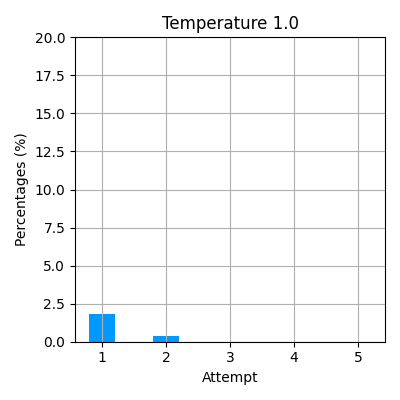}
        \caption{Reverse History.}
    \end{subfigure}
    \caption{Successful verifications by attempt for temperature 1.0.}
    \label{fig:esbmc_ai_successful_by_attempt_per_temp_1.0}
\end{figure}

\begin{figure}[H]
    \centering
    \begin{subfigure}[t]{0.32\textwidth}
        \centering
        \includegraphics[width=\textwidth]{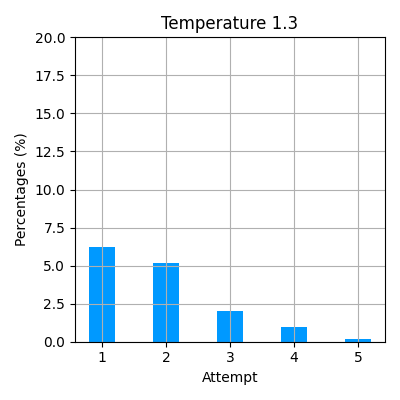}
        \caption{Latest State Only.}
    \end{subfigure}
    \hfill
    \begin{subfigure}[t]{0.32\textwidth}
        \centering
        \includegraphics[width=\textwidth]{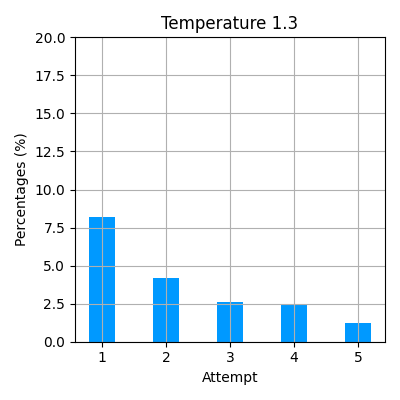}
        \caption{Forward History.}
    \end{subfigure}
    \hfill
    \begin{subfigure}[t]{0.32\textwidth}
        \centering
        \includegraphics[width=\textwidth]{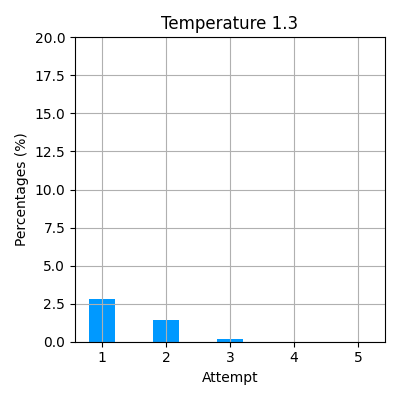}
        \caption{Reverse History.}
    \end{subfigure}
    \caption{Successful verifications by attempt for temperature 1.3.}
    \label{fig:esbmc_ai_successful_by_attempt_per_temp_1.3}
\end{figure}

\cref{fig:esbmc_ai_successful_by_attempt_per_temp_0.0,fig:esbmc_ai_successful_by_attempt_per_temp_0.4,fig:esbmc_ai_successful_by_attempt_per_temp_0.7,fig:esbmc_ai_successful_by_attempt_per_temp_1.0,fig:esbmc_ai_successful_by_attempt_per_temp_1.3} show the next set of experiments, they are to find out the successful verifications by attempt for each temperature tested. For the LSO experiments, as the temperature is increased from 0, the number of successful repairs at each attempt does not change much. The only exception is temperature 1.0, where the performance is significantly lower for attempts 0 and 1. Furthermore, the Forward History experiments only decrease in successful repairs on the attempt 0, as the temperature is increased. The other attempts are largely unaffected. The only exception to this is on temperature 0.4, however, this fluctuation can be considered as experimental noise. The Reverse History experiments had only a couple of successful repairs in attempt 0 on temperature 0.7 and higher. All other attempts after the 2nd attempt yield 0 successful repairs.

\begin{tcolorbox}
\textbf{Changing the Temperature Takeaway.} Increasing the temperature for Forward History experiments only affects the 1st attempt. The LSO experiments did not noticeably change in the amount of successful repairs as the temperature is varied. 
\end{tcolorbox}

\subsection{Successful Verifications By Attempt Per Prompt (Temperature 0.0)}

\begin{figure}[H]
    \centering
    \begin{subfigure}[t]{0.32\textwidth}
        \centering
        \includegraphics[width=\textwidth]{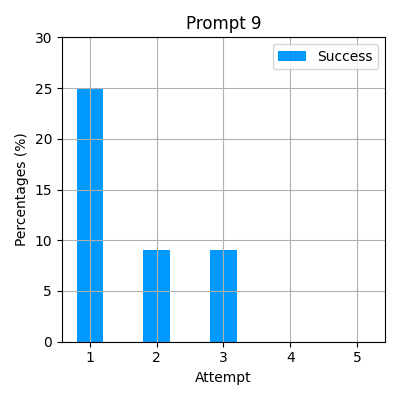}
        \caption{Forward History.}
    \end{subfigure}
    \caption{Successful verifications by attempt for prompt 9 (temperature 0.0).}
    \label{fig:esbmc_ai_successful_by_attempt_per_prompt_0}
\end{figure}

\begin{figure}[H]
    \centering
    \begin{subfigure}[t]{0.32\textwidth}
        \centering
        \includegraphics[width=\textwidth]{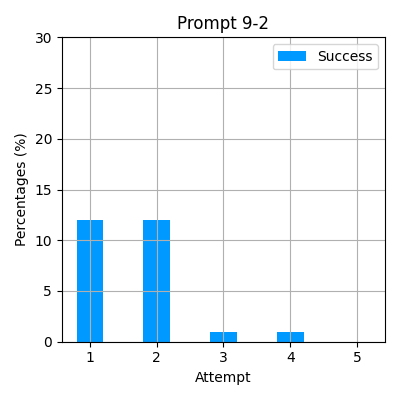}
        \caption{Latest State Only.}
    \end{subfigure}
    \begin{subfigure}[t]{0.32\textwidth}
        \centering
        \includegraphics[width=\textwidth]{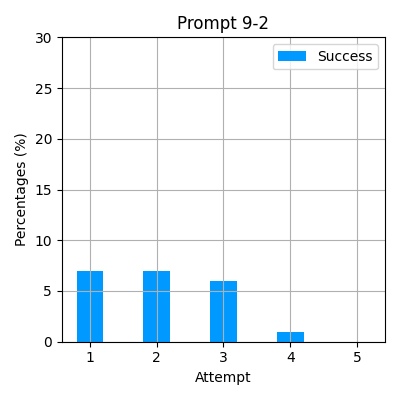}
        \caption{Forward History.}
    \end{subfigure}
    \hfill
    \caption{Successful verifications by attempt for prompt 9-2 (temperature 0.0).}
    \label{fig:esbmc_ai_successful_by_attempt_per_prompt_1}
\end{figure}

\begin{figure}[H]
    \centering
    \begin{subfigure}[t]{0.32\textwidth}
        \centering
        \includegraphics[width=\textwidth]{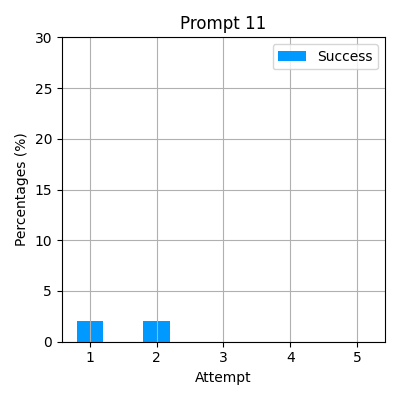}
        \caption{Latest State Only.}
    \end{subfigure}
    \begin{subfigure}[t]{0.32\textwidth}
        \centering
        \includegraphics[width=\textwidth]{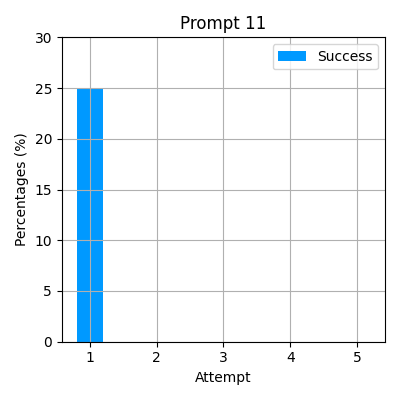}
        \caption{Forward History.}
    \end{subfigure}
    \hfill
    \caption{Successful verifications by attempt for prompt 11 (temperature 0.0).}
    \label{fig:esbmc_ai_successful_by_attempt_per_prompt_2}
\end{figure}

\begin{figure}[H]
    \centering
    \begin{subfigure}[t]{0.32\textwidth}
        \centering
        \includegraphics[width=\textwidth]{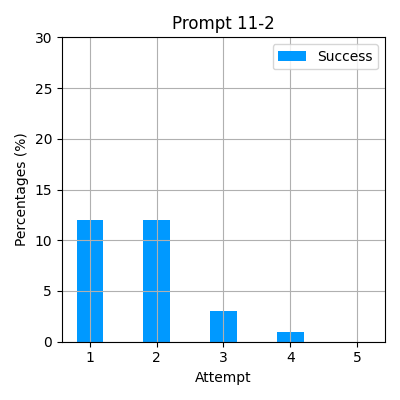}
        \caption{Latest State Only.}
    \end{subfigure}
    \begin{subfigure}[t]{0.32\textwidth}
        \centering
        \includegraphics[width=\textwidth]{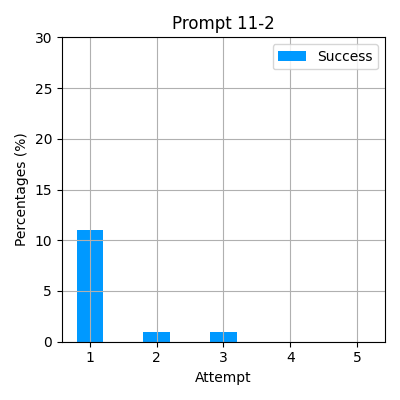}
        \caption{Forward History.}
    \end{subfigure}
    \hfill
    \caption{Successful verifications by attempt for prompt 11-2 (temperature 0.0).}
    \label{fig:esbmc_ai_successful_by_attempt_per_prompt_3}
\end{figure}

At this stage, it is obvious that the best performing configuration for repairing AI C code is using temperature 0.0, and Forward History. \cref{fig:esbmc_ai_successful_by_attempt_per_prompt_0,fig:esbmc_ai_successful_by_attempt_per_prompt_1,fig:esbmc_ai_successful_by_attempt_per_prompt_2,fig:esbmc_ai_successful_by_attempt_per_prompt_3} show the percentage of successful repairs at each attempt for each prompt. As seen from the previous figures, the Reverse History and LSO experiments did not have any successful repairs for the prompt 9, so there is no figure shown for those experiments. Also, the old ESBMC-AI prompt did not successfully repair a single sample for temperature 0.0, so it is not shown either. The remaining bar graphs for the LSO experiments show that the most successful repairs occurred in prompt 9-2 and prompt 11-2, where the same amount of samples were repaired in the first two attempts.

For the Forward History experiments, prompt 9 has the highest performance, showing that samples were repaired in the first 3 attempts, after attempt 2, no further samples were repaired. Prompt 9-2 had a much lower number of successful repairs per attempt, however, there were some successful repairs on attempt 3. Prompt 11 had no successful repairs of samples during any retries, and only repaired samples successfully on the first attempt.

\begin{tcolorbox}
\textbf{Takeaway.} For AI C code one line repair, the optimal total number of attempts is 3.
\end{tcolorbox}

\end{document}